\documentclass[aps,prb,twocolumn,amsmath,amssymb,nofootinbib,
  superscriptaddress]{revtex4-2}

\usepackage{graphicx}
\usepackage{bm}
\usepackage{amsmath}
\usepackage{amssymb}
\usepackage{physics}
\usepackage[colorlinks=true,linkcolor=blue,urlcolor=blue,citecolor=blue]{hyperref}
\usepackage[usenames,svgnames]{xcolor}
\usepackage{makecell}
\usepackage[normalem]{ulem}

\newcommand{\Hop}{\hat{H}}
\newcommand{\Heff}{H_{\rm eff}}

\newcommand{\aop}{\hat{a}}
\newcommand{\adop}{\hat{a}^{\dagger}}

\newcommand{\nudt}{College of Science, National University of Defense Technology, Changsha 410073, China}

\newcommand{\ustc}{State Key Laboratory of Precision and Intelligent Chemistry, University of Science and Technology of China, Hefei 230026, China}
\newcommand{\hefeilab}{Hefei National Laboratory, University of Science and Technology of China, Hefei 230088, China}

\begin{document}

\title{Clifford augmented density matrix renormalization group for \textit{ab initio} quantum chemistry}

\author{Lizhong Fu}
\affiliation{\ustc}

\author{Honghui Shang}
\email{shanghui.ustc@gmail.com}
\affiliation{\ustc}

\author{Jinlong Yang}
\email{jlyang@ustc.edu.cn}
\affiliation{\ustc}
\affiliation{\hefeilab}

\author{Chu Guo}
\email{guochu604b@gmail.com}
\affiliation{\nudt}



\begin{abstract}
The recently proposed Clifford augmented density matrix renormalization group (CA-DMRG) method seamlessly integrates Clifford circuits with matrix product states, and takes advantage of the expression power from both. 
CA-DMRG has been shown to be able to achieve higher accuracy than standard DMRG on commonly used lattice models, with only moderate computational overhead compared to the latter.  In this work, we propose an efficient scheme in CA-DMRG to deal with \textit{ab initio} quantum chemistry Hamiltonians, and apply it to study several molecular systems. Our numerical results show that CA-DMRG can reach higher accuracy than DMRG using the same bond dimension, pointing out a promising route to push the boundary of solving \textit{ab initio} quantum chemistry with strong static correlations.
\end{abstract}

\maketitle


\section{Introduction}
A major challenge of \textit{ab initio} quantum chemistry is to solve the many-electron Schr\"odinger equation to find the ground state. An exact treatment is generally hard due to the exponential growth of the Hilbert space size. Therefore, the brute-force approach, such as the full configuration interaction (FCI) method, is limited to small system sizes~\cite{VogiatzisJong2017,GaoYoshida2024}. A large number of approximate but more scalable methods have been developed. Among these, the Hartree-Fock theory~\cite{Hartree1928,Fock1930}, density functional theory~\cite{KohnSham1965}, truncated configuration interaction~\cite{HolmesUmrigar2016,SchriberEvangelista2017} and the coupled cluster theory~\cite{CCSD-Purvis1982AFC} could often scale up to large systems with thousands of electrons or beyond, but may not be reliable for cases with strong static correlations. The quantum Monte Carlo method~\cite{FoulkesRajagopal2001} and DMRG~\cite{White1992,White1993}, instead, could obtain accurate results in strongly correlated cases, but could suffer from the sign problem or only be efficient for relatively small systems (usually less than $100$ electrons). 
The recently proposed neural network quantum state method~\cite{CarleoTroyer2017}, which is a combination of variational quantum Monte Carlo and neural network ansatz, has also been applied for quantum chemistry calculations, which seems promising as an accurate and scalable method~\cite{ChooCarleo2020,HermannNoe2020,PfauFoulkes2020,BarrettLvovsky2022,WuShang2023,VonPfau2022}. However, its accuracy for large-scale calculations is yet to be demonstrated.

DMRG plays a very special rule among its alternatives. It is particularly useful for molecular systems with strong static correlations, as it can usually deliver  reliable results given enough computational resources~\cite{Xiang1996,WhiteMartin1999,MartiReiher2008,ChanSharma2011,KurashigeYanai2013,WoutersNeck2014,LiChan2019,LarssonChan2022,XiangLi2023,Guo2022b}. DMRG parameterizes the wave function as a special one-dimensional tensor network: matrix product state (MPS). The overall accuracy and efficiency of DMRG are essentially determined by a single integer hyperparameter, which is referred to as the \textit{bond dimension} and denoted as $\chi$, and its computational cost grows as $O(\chi^3)$. A crucial advantage of DMRG is that it can often reach the global minimum and its accuracy increases monotonically with $\chi$, therefore one could easily perform extrapolations to extract the asymptotic ground state energy.

However, DMRG is limited by the expressive power of MPS. It is known that MPS can only generate quantum states with low entanglement,
and is less expressive compared to most other commonly used tensor network ansatz or neural networks~\cite{DengSarma2017,SharirCarleo2022,WuCarleo2023}.
A recent effort inspired from quantum computing and quantum information is to enhance the expressive power of MPS by applying a Clifford circuit on top of it, referred to as the Clifford augmented MPS (CAMPS)~\cite{MishmashMezzacapo2023,LlimaSaez2024,LamiNardis2025}. In contrary to MPS, the Clifford circuit can easily generate an arbitrary amount of entanglement while remaining efficiently classically simulable~\cite{Gottesman1997,AaronsonGottesman2004,AndersBriegel2006}. 
Therefore CAMPS can in principle represent quantum states with volume-law entanglement, 
hopefully being more powerful than MPS or Clifford circuit alone. 
The recently proposed CA-DMRG algorithm makes full use of the CAMPS ansatz: by slightly adjusting the standard DMRG algorithm, it can optimize the underlying MPS and Clifford circuit at the same time, with only moderate computational overhead compared to standard DMRG~\cite{QianQin2024}. Till now CA-DMRG has been shown to be able to achieve higher accuracy than DMRG on two-dimensional~\cite{QianQin2024} or critical one-dimension lattice models~\cite{FanXiang2025}.


In this work, we aim to apply CA-DMRG for \textit{ab initio} quantum chemistry calculations. On the technical side, we propose an efficient scheme to update the matrix product operator (MPO) representation of the Hamiltonian during CA-DMRG to reduce the computational overhead for quantum chemistry Hamiltonians. We apply the adjusted CA-DMRG algorithm to calculate the ground states for a number of molecular systems, including H$_2$O, NH$_3$, C$_2$ and N$_2$, and to study the potential energy curve of the N$_2$ molecular system. Our numerical results show that with CA-DMRG we can reach higher accuracy than standard DMRG using the same bond dimension, demonstrating that CA-DMRG is a promising tool for accurate solutions of \textit{ab initio} quantum chemistry.

\begin{figure*}
    \centering
    \includegraphics[width=\linewidth]{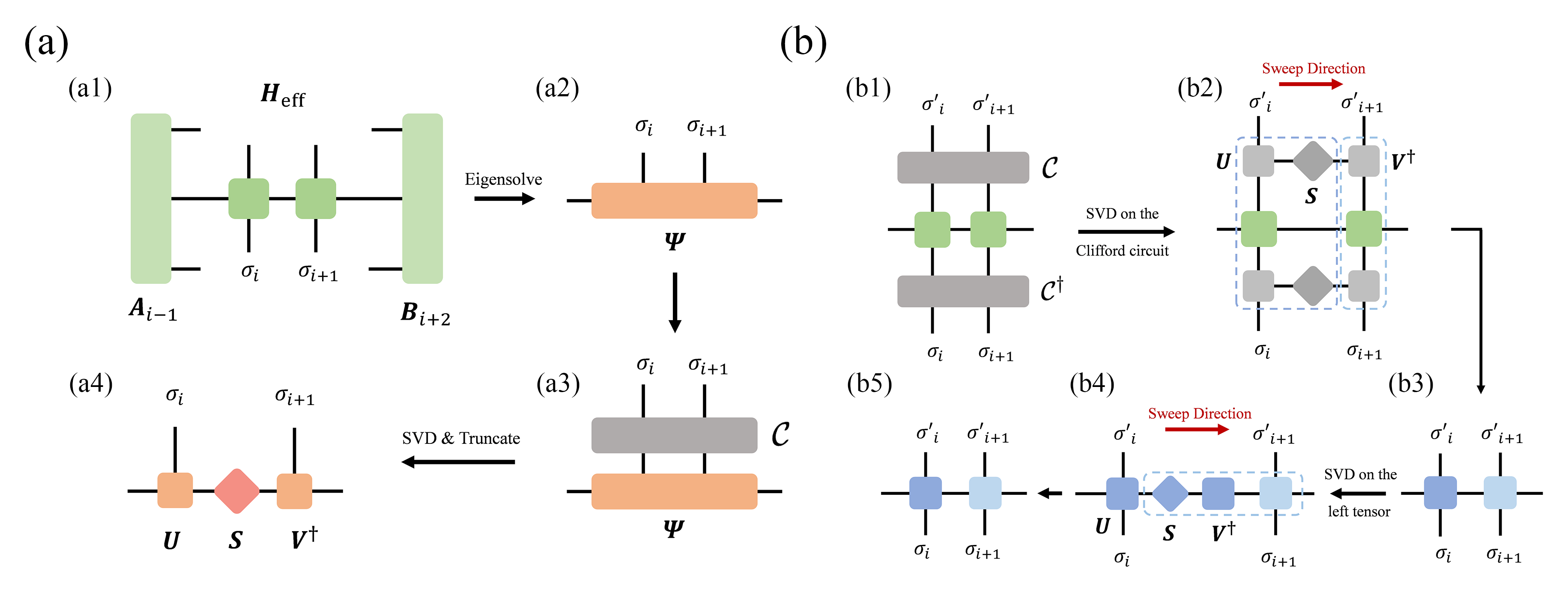}
    \caption{(a) Local minimization and SVD truncation in CA-DMRG, which contains 4 major steps: (a1) building the two-site effective Hamiltonian $\Heff$; (a2) performing eigenvalue decomposition to obtain the lowest eigenvalue and eigenvector $\Psi$ of $\Heff$; (a3) selecting a two-site Clifford circuit and apply it onto $\Psi$; (a4) performing SVD truncation on the resulting tensor. Steps (a3, a4) are repeated for all the possible two-site Clifford circuits to obtain the optimal Clifford circuit which minimizes the thrown weight during SVD truncation. Only steps (a1, a2, a4) are required in standard two-site DMRG. (b) Updating the local MPO tensors according to $\mathcal{C}\Hop\mathcal{C}^{\dagger}$ after one obtains the optimal two-site Clifford circuit $\mathcal{C}$ from step (a), which is done in $5$ steps as shown in panels (b1-b5). The dashed blue box in (b2) and (b4) means that the tensors inside are contracted first.}
    \label{fig:demo}
\end{figure*}

\section{Method}
\label{sec:Method}

In this section, we first briefly review DMRG and CA-DMRG, then we present our scheme to handle the MPO representation of the Hamiltonian in CA-DMRG which is designed for quantum chemistry Hamiltonians.

\subsection{Standard DMRG}

DMRG can be most conveniently presented in the language of MPS and MPO~\cite{Schollwock2011,ChanWhite2016}. 
In this work we assume that the wave function is written in the basis of spin orbitals, such that the local Hilbert space has $2$ degrees of freedom, denoted as $\{\sigma=0,1\}$. For a molecular system with $N$ spin orbitals, the wave function $\vert \psi\rangle$ can be generally written as a rank-$N$ tensor, denoted as $c^{\sigma_1\sigma_2\cdots\sigma_N}$, which contains $2^N$ scalars. The Hamiltonian $\Hop$ can be generally written as a rank-$2N$ tensor, denoted as $h_{\sigma_1\sigma_2\cdots\sigma_N}^{\sigma_1'\sigma_2'\cdots\sigma_N'}$. In DMRG, the wave function is represented as an MPS:
\begin{align}
    c^{\sigma_1\sigma_2\cdots\sigma_N} = \sum_{\{\alpha_i\}}M_{\alpha_0, \alpha_1}^{\sigma_1}M_{\alpha_1, \alpha_2}^{\sigma_2}\cdots M_{\alpha_{N-1}, \alpha_N}^{\sigma_N},
\end{align}
where $\alpha_i$ represents the auxiliary degree of freedom, $\alpha_0, \alpha_N$ are trivial indices of size $1$, which are added for notational convenience, $M_{\alpha_{i-1}, \alpha_i}^{\sigma_i}$ represents the $i$th site tensor of MPS, which has one physical index plus two auxiliary indices. 
The size of the largest auxiliary index is referred to as the bond dimension, denoted as $\chi = \max_{1\leq i< N} \left(\dim(\alpha_i)\right)$.
Similarly, the Hamiltonian can be represented as an MPO:
\begin{align}
h_{\sigma_1\sigma_2\cdots\sigma_N}^{\sigma_1'\sigma_2'\cdots\sigma_N'} = \sum_{\{\beta_i\}}W_{\beta_0, \beta_1}^{\sigma_1, \sigma_1'}W_{\beta_1, \beta_2}^{\sigma_2, \sigma_2'}\cdots W_{\beta_{N-1}, \beta_N}^{\sigma_N, \sigma_N'},
\end{align}
where $\beta_i$ represents the auxiliary degree of freedom, $\beta_0, \beta_N$ are trivial indices, $W_{\beta_{i-1}, \beta_i}^{\sigma_i, \sigma_i'}$ represents the $i$th site tensor of MPO, which has two physical indices plus two auxiliary indices. The bond dimension of an MPO is denoted as $\chi_w = \max_{1\leq i< N} \left(\dim(\beta_i)\right)$. The MPO representation of the Hamiltonian is a given input of DMRG. In the single-site version of DMRG, the MPS ansatz of the wave function is variationally optimized using the following strategy: the MPS is kept in the mixed-canonical form with the site tensor to be optimized as the canonical center (for the canonical form and more detailed introduction of the standard DMRG algorithm, one could refer to excellent reviews such as Refs.~\cite{Schollwock2011,Orus2014}), the site tensors are optimized one by one from left to right and then from right to left, and one such process is referred to as a \textit{sweep}. The two-site version of DMRG is similar to the single-site DMRG, but with two consecutive site tensors optimized in each local minimization step. In practice the two-site DMRG is often preferred, especially when quantum symmetries are considered, as it is less prone to be trapped in local minima, and it can dynamically adjust the bond dimension.

\subsection{Clifford augmented DMRG}
CA-DMRG uses CAMPS as the underlying parametric ansatz, in which not only the MPS ansatz, but also the Clifford circuit is variationally optimized during the sweep.
The CAMPS ansatz can be formally written as $\mathcal{C}\vert {\rm MPS}\rangle$, where $\mathcal{C}$ denotes the Clifford circuit and $\vert {\rm MPS}\rangle$ denotes the MPS. 
The idea of CA-DMRG to achieve simultaneous optimization of both $\mathcal{C}$ and $\vert {\rm MPS}\rangle$ is simple but elegant: (i) once obtaining the two-site tensor as the local ground state, e.g., the rank-$4$ tensor $\Psi$ in Fig.~\ref{fig:demo}(a2), one selects a two-site Clifford circuit and applies it onto $\Psi$ as in Fig.~\ref{fig:demo}(a3), then one performs singular value decomposition (SVD) on the resulting tensor and truncate the small singular values (SVD truncation), while in the standard two-site DMRG, one directly performs SVD truncation on $\Psi$; (ii) after that, one transforms the Hamiltonian as $\mathcal{C}\Hop\mathcal{C}^{\dagger}$. 
For model Hamiltonians with short-ranged interactions,
step (i) is the place where the computational overhead of CA-DMRG compared to the standard DMRG occurs: one needs to try all the possible two-site Clifford circuits (and there are $720$ of them) to find the optimal one which minimizes the thrown weight during SVD truncation, therefore one needs to do $720$ SVDs instead of a single SVD in standard DMRG (here we can also see that CA-DMRG is at least as accurate as DMRG, as one could apply the trivial identity circuit in Fig.~\ref{fig:demo}(a3), which will correspond to DMRG). 
In step (ii), as the Clifford circuit preserves the number of Pauli strings, the total number of Pauli strings in the Hamiltonian keeps the same, and if the number is not large, the cost of this step is negligible. However, as we will show in the following, this is not the case for quantum chemistry Hamiltonians.

\subsection{CA-DMRG for quantum chemistry Hamiltonians}\label{sec:qccamps}

The electronic Hamiltonian of a chemical system can be written in the second-quantized formalism:
\begin{align}\label{eq:ham}
\Hop = \sum_{pq}t_{pq}\adop_p\aop_q + \frac{1}{2}\sum_{pqrs}g_{pqrs}\adop_p\adop_q\aop_r\aop_s,
\end{align}
where $p,q,r,s$ are orbital labels (the spin label $\sigma$ is omitted), $\adop$ and $\aop$ are fermionic creation and annihilation operators, $t_{pq}$ and $g_{pqrs}$ are one- and two-electron integrals in molecular orbital basis. 

In a local minimization step of DMRG or CA-DMRG, the two computationally expensive operations are (1) performing the eigenvalue decomposition of $\Heff$ and (2) performing SVD decomposition on the two-site tensor, the costs of these two steps are $O(\chi^3\chi_w + \chi^2\chi_w^2)$ and $O(\chi^3)$ respectively. For model Hamiltonians, the total number of Pauli strings usually scale as $O(N)$, and the bond dimension $\chi_w$ scale at most as $O(N)$, as such one often has $\chi_w \ll \chi$ and these two steps have a comparable cost (in fact, the SVD operation could be more expensive as it has a larger prefactor).

However, for quantum chemistry Hamiltonians as in Eq.(\ref{eq:ham}),
the total number of Pauli strings scales as $O(N^4)$ after the Jordan-Wigner transformation~\cite{JordanWigner1928}. When written as MPO, the bond dimension $\chi_w$ would scale as $O(N^2)$ at best~\cite{SharmaChan2012}. Therefore in such cases, the cost of (1) is generally much larger than the cost of (2). 
Even worse, the Clifford transformation does not necessarily preserve $\chi_w$. With a large $\chi_w$, building the MPO representation of the transformed Hamiltonian $\mathcal{C}\Hop\mathcal{C}^{\dagger}$ from beginning could itself be expensive, particularly if this needs to be done after each local minimization. To maximally reuse the MPO representation, we propose a local updating scheme for $\Hop$ after each local minimization step, which is illustrated in Fig.~\ref{fig:demo}(b). The idea is to keep the MPO in the mixed-canonical form with the same canonical center as the underlying MPS, and then transform the canonical center with the obtained optimal two-site Clifford circuit only. Similar to the SVD truncation of MPS, the computational cost of this operation scales as $O(\chi_w^3)$. The step in Fig.~\ref{fig:demo}(b2) is merely used to improve efficiency: adding this step would be slightly more advantageous if the Clifford circuit $\mathcal{C}$ is not full-rank, which is usually the case (if we encounter several Clifford circuits which are all optimal, we will choose the one with smallest rank).

To this end, we analyze the computational overhead of CA-DMRG compared to standard DMRG in detail. First, in terms of the number of parameters, CAMPS adds a Clifford circuit on top of the MPS ansatz. However, this information can be stored at negligible memory cost, e.g., one could store each Clifford gate as a symbol and for $m$ sweeps there are only $2m N$ symbols to be stored. In fact, if one is not interested in the ground state wave function but only cares about the ground state energy, the information of the optimized Clifford circuit does not need to be stored at all. In terms of the computational cost, one needs to perform $720$ SVDs on the locally optimized two-site MPS tensor, each with cost $O(\chi^3)$, and to perform an additional SVD on the MPO tensor, with cost $O(\chi_w^3)$, per local minimization step. Assuming that $\chi_w$ remains more or less the same during CA-DMRG (which is roughly the case, as illustrated in our numerical examples), the latter operation adds a cost of $O(N^6)$. In addition, it has been shown that for the quantum chemistry Hamiltonian, one could explore the sparsity of the MPO to further reduce the computational cost of the local minimization step from $O(\chi^3 N^2 + \chi^2N^4)$ to $O(\chi^3 N^2 + \chi^2N^3)$. In comparison, in our CA-DMRG algorithm, the sparsity of the initial MPO will be destroyed by SVD. 

We also note that truncation of small singular values is necessary after applying the two-site Clifford circuit onto the local MPO tensors, otherwise $\chi_w$ will grow exponentially with sweeps. However, it is a known issue that SVD truncation of MPO could suffer from numerical inaccuracy for large scale problems, for example if there is a large constant term in $\Hop$. In our work we use a SVD truncation threshold of $10^{-10}$, e.g., we throw away the singular values if their relative weights are smaller than the threshold. In our numerical test, we did not encounter such a numerical issue as the number of considered spin orbitals is not too large. More stable compression schemes for MPO may be explored in future large-scale investigations~\cite{HubigSchollwock2017,WhiteRefael2018,RoySlagle2024}.

\section{Numerical results}

\begin{figure}
    \centering
    \includegraphics[width=\linewidth]{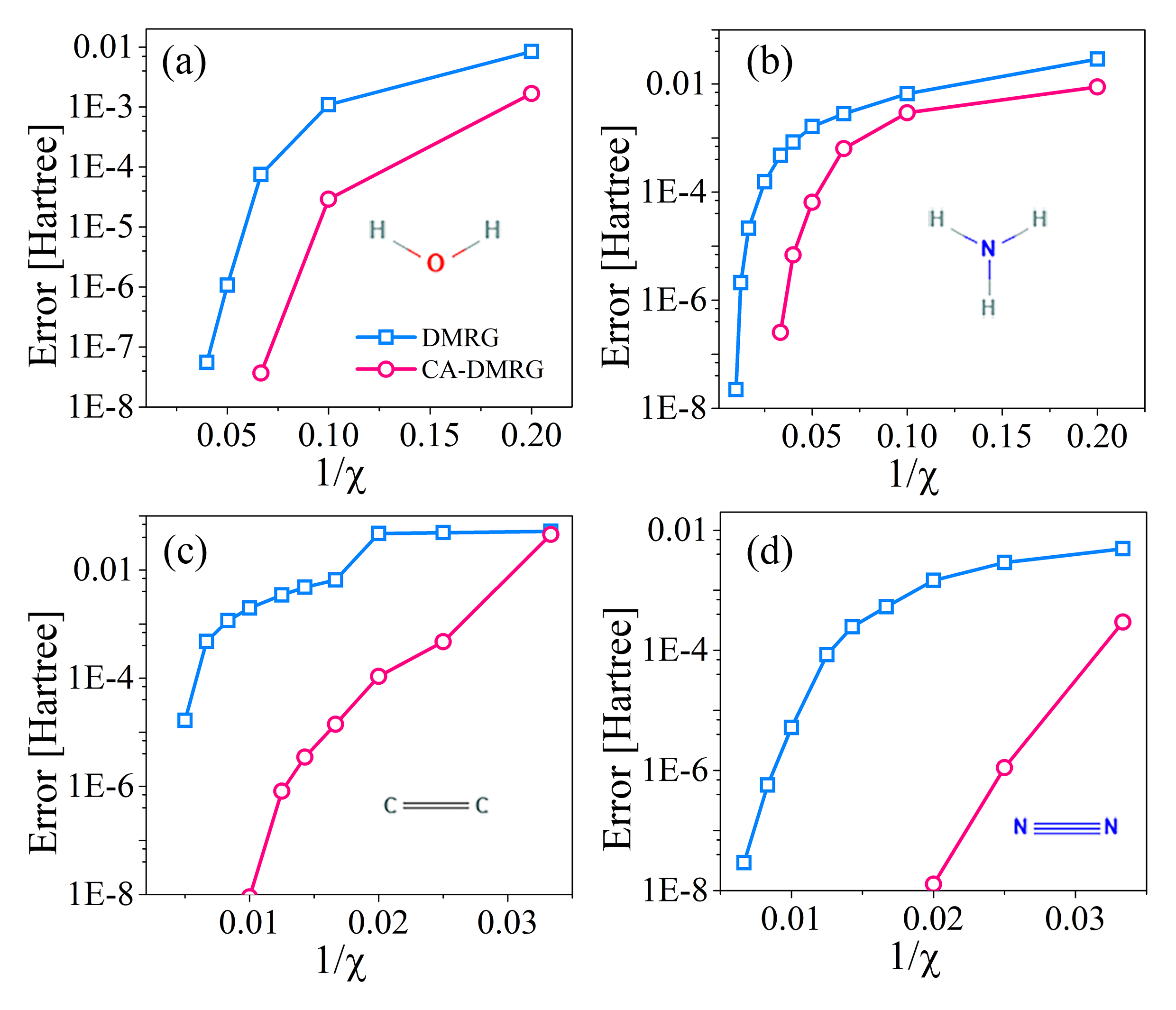}
    \caption{The errors between the ground state energies and the corresponding FCI energies for (a) H$_2$O, (b) NH$_3$, (c) C$_2$ and (d) N$_2$ as functions of the inverse bond dimension. The CA-DMRG and two-site DMRG results are represented by the red line with circle and by the blue line with square respectively. We have used the STO-3G basis set for all the molecules. }
    \label{fig:energy}
\end{figure}

In the following we apply our adapted CA-DMRG to study a variety of molecular systems, with up to $20$ spin orbitals. Similar to Refs.~\cite{QianQin2024,FanXiang2025}, we will compare the accuracy of our simulations to standard two-site DMRG calculations under the same bond dimension. For both DMRG and CA-DMRG, we start from random initial guesses and use at most $50$ sweeps. We will terminate the program in advance if the number of sweeps is larger than $20$ and the absolute error between two successive sweeps is smaller than $10^{-6}$.  In cases where a single run of DMRG or CA-DMRG is trapped in local minima, we will run $5$ instances and choose the one with the lowest energy, as a simple way to avoid local minima (the ground state converged by DMRG could also be a good initial guess for CA-DMRG in real applications).

We first test our method on several commonly used small-scale molecular systems, namely H$_2$O, NH$_3$, C$_2$ and N$_2$ in the STO-3G basis set, which contain $14$, $18$, $20$ and $20$ spin orbitals respectively. 
We calculated the ground state energy using both CA-DMRG and two-site DMRG with different bond dimensions, and the results are shown in Fig.~\ref{fig:energy}. 
The FCI energy in the same basis set is used as reference. 
We can see that with CA-DMRG we can always obtain lower energy than DMRG, generally by one to two orders of magnitude. 
In particular, CA-DMRG already reaches chemical accuracy ($1.6\times 10^{-3}$ Hartree) with $\chi=40$ for the C$_2$ molecule, while two-site DMRG needs $\chi=100$ to reach the same accuracy.

\begin{figure}
    \centering
    \includegraphics[width=\linewidth]{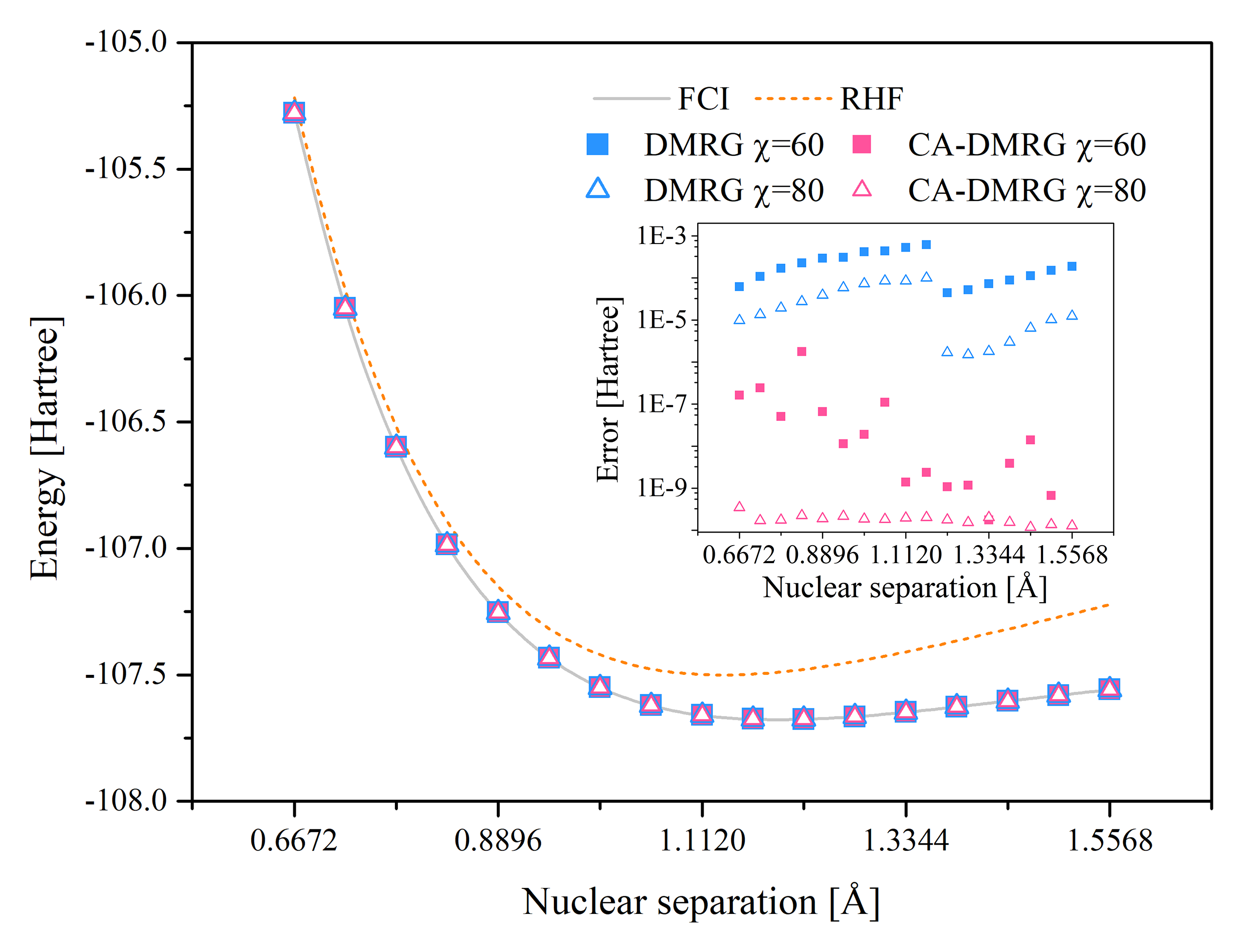}
    \caption{Potential energy curve calculated for the N$_2$ molecular system in the STO-3G basis set. The blue and red squares represent two-site DMRG and CA-DMRG results calculated with $\chi=60$, while the blue and red empty triangles represent DMRG and CA-DMRG results calculated with $\chi=80$. The restricted Hartree Fock (RHF, the orange dashed line) and FCI (the gray solid line) energies are also shown as reference. The inset shows the errors against the corresponding FCI energies.  }
    \label{fig:eos}
\end{figure}

In Fig.~\ref{fig:eos}, we calculate the potential energy curve of the N$_2$ molecule at different bond dimensions, using both CA-DMRG and two-site DMRG. The inset shows the errors against the FCI energies. We can see that
the CA-DMRG results are more accurate than the two-site DMRG results for all the different molecular bond lengths considered, by generally $2$ orders of magnitude (the difference is about $4$ orders of magnitude for $\chi=80$). These results demonstrate the superior accuracy of CA-DMRG across different interaction strengths. 


\begin{figure}
    \centering
    \includegraphics[width=0.8\linewidth]{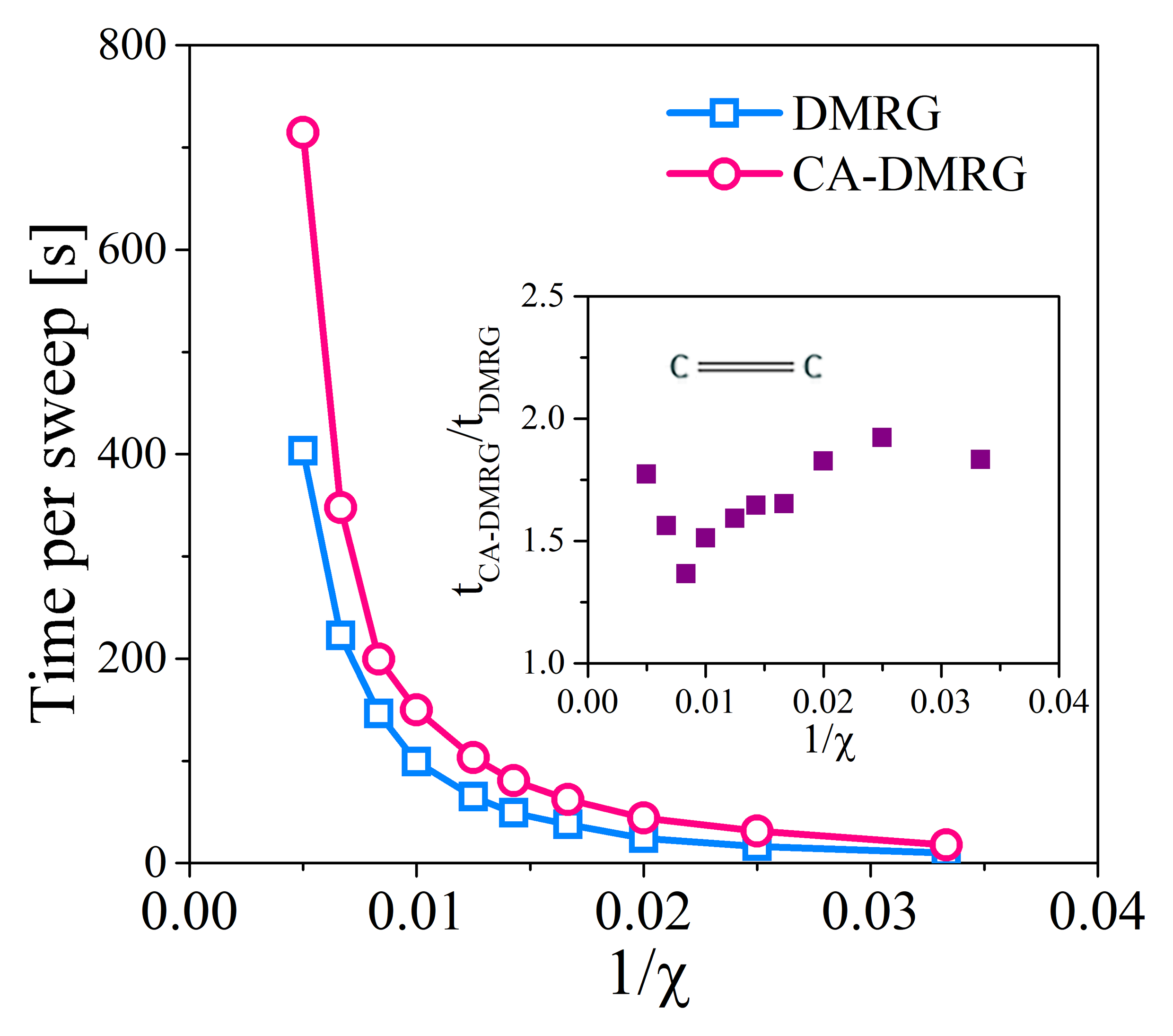}
    \caption{The run time per sweep for the C$_2$ molecule as a function of the inverse bond dimension. The CA-DMRG and two-site DMRG results are represented by the red line with empty circle and by the blue line with empty square respectively. The purple squares in the inset show the ratio between the run time of CA-DMRG and that of DMRG as a function of the inverse bond dimension.}  
    \label{fig:runtime}
\end{figure}

Next, we analyze the numerical performance of CA-DMRG from different perspectives. 
In Fig.~\ref{fig:runtime}, we show a comparison of the run time per sweep between CA-DMRG and DMRG, as a function of the inverse bond dimension, for the C$_2$ molecule (which is the hardest one to converge for both methods from Fig.~\ref{fig:energy}). We can see that DMRG is faster than CA-DMRG by roughly a constant factor ($1.2\sim 2$) for all the different bond dimensions considered. However, here we stress that our implementation of DMRG is far from optimal: both the quantum symmetries and the sparsity of the MPO are not explored.

\begin{figure}
    \centering
    \includegraphics[width=\linewidth]{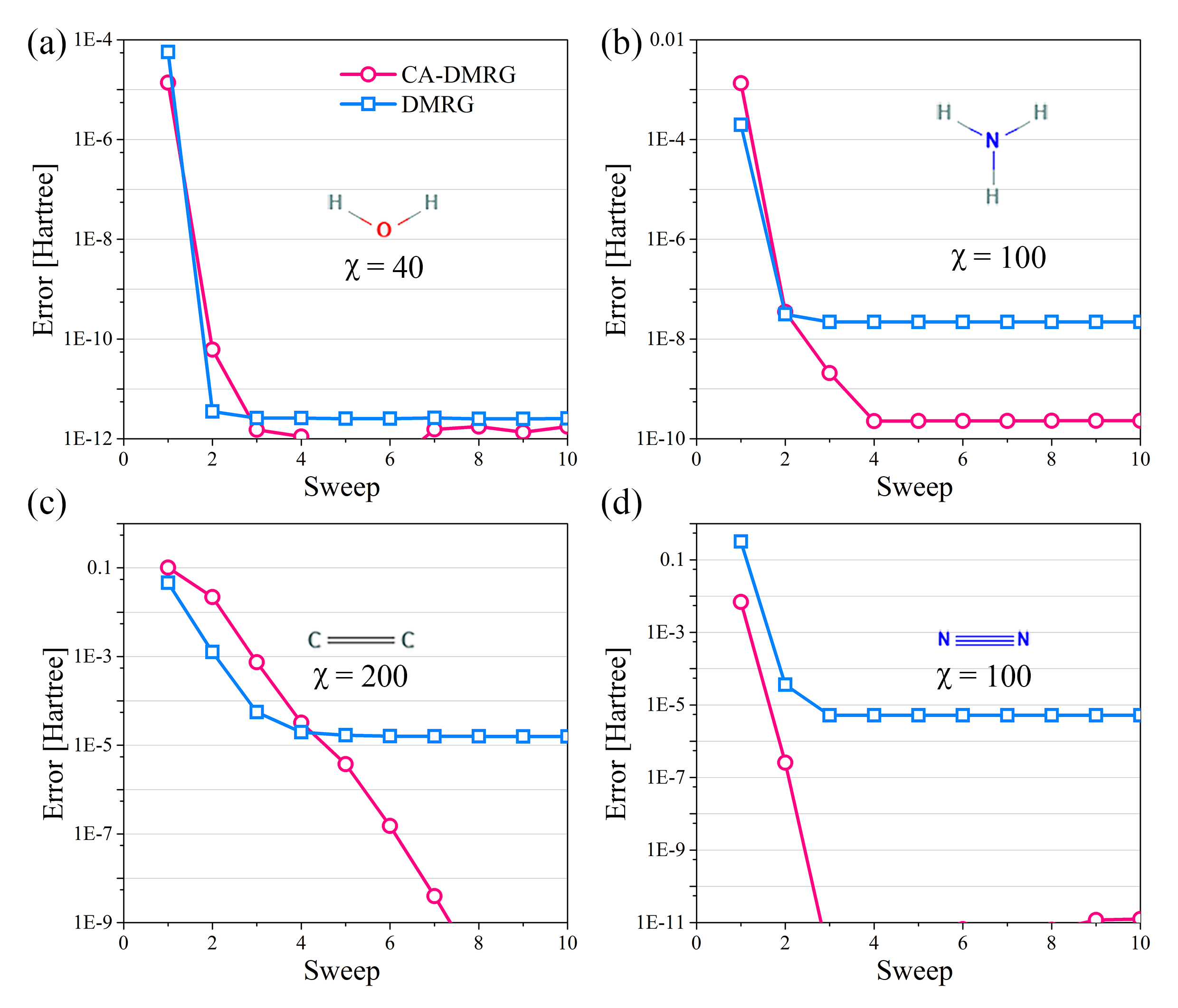}
    \caption{The errors between the ground state energies and the corresponding FCI energies for (a) H$_2$O with $\chi=40$, (b) NH$_3$ with $\chi=100$, (c) C$_2$ with $\chi=200$ and (d) N$_2$ with $\chi=100$ as functions of the number of sweeps. The CA-DMRG and DMRG results are represented by the red line with circle and by the blue line with square respectively.
    }
    \label{fig:error}
\end{figure}

In Fig.~\ref{fig:error}, we show the convergence of CA-DMRG and DMRG as functions of the number of sweeps for the four molecular systems used in Fig.~\ref{fig:energy}, for which we have chosen bond dimensions $\chi=40, 100, 200, 100$ respectively such that chemical accuracies have been well reached. We can see that DMRG converges within $2$ to $4$ sweeps for these four molecular systems. CA-DMRG requires a slightly larger number of sweeps to converge, but can reach much accuracy, especially for the C$_2$ and N$_2$ molecular systems.

\begin{figure}
    \centering
    \includegraphics[width=\linewidth]{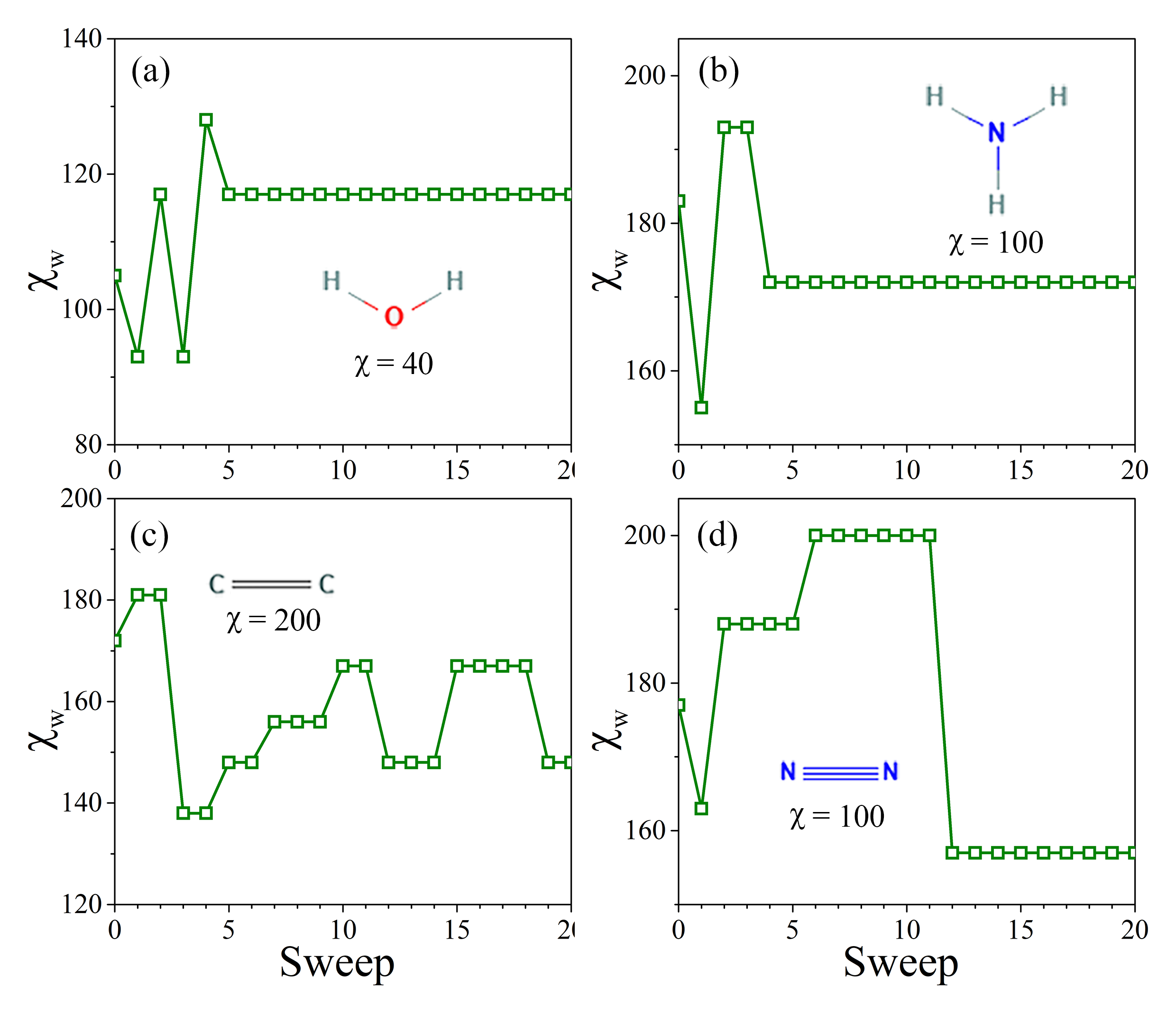}
    \caption{The bond dimension of MPO, $\chi_w$, as a function of sweep during the CA-DMRG calculation for (a) H$_2$O with $\chi=40$, (b) NH$_3$ with $\chi=100$, (c) C$_2$ with $\chi=200$ and (d) N$_2$ with $\chi=100$. The CA-DMRG calculations used here are taken from Fig.~\ref{fig:energy} with particular choices of the bond dimensions for the MPS.}
    \label{fig:bonddim}
\end{figure}

As we have discussed in Sec.~\ref{sec:qccamps}, for quantum chemistry Hamiltonian the cost of transforming the Hamiltonian, e.g., the operation $\mathcal{C}\Hop\mathcal{C}^{\dagger}$, is no longer negligible. In our scheme, we build the MPO representation of $\Hop$ and then transform the local MPO tensors after each local minimization step. During this process, the bond dimension $\chi_w$ of the MPO could grow with sweeps in principle. 
In fact, in the worst case $\chi_w$ could scale as $O(N^4)$, the same as the total number of Pauli strings in $\Hop$ (in the most naive approach to build the MPO representation of a Hamiltonian, $\chi_w$ would be the same as the number of Pauli strings~\cite{HubigSchollwock2017}).
In Fig.~\ref{fig:bonddim}, we plot $\chi_w$ as a function of sweeps, in one to one correspondence with the four molecular systems considered in Fig.~\ref{fig:energy}. The initial MPO is constructed using the algorithm in Ref.~\cite{ChanWhite2016}, and then prepared in the right-canonical form using a right to left sweep (this sweep is merely for the purpose of CA-DMRG, and will not change the bond dimension in general).
We can see that although being operated on by two-site Clifford circuits
very frequently, $\chi_w$ has neither been growing exponentially, nor growing towards the $O(N^4)$ limit. Instead, it only fluctuates mildly in the first few sweeps and then remained mostly unchanged. Interestingly, for the NH$_3$, C$_2$ and N$_2$ molecular systems, the bond dimensions in the last few sweeps are even smaller than the bond dimension of the initial MPO.
This result implies that the computational cost for our adapted CA-DMRG algorithm will still remain comparable to standard DMRG.

\begin{figure}
    \centering
    \includegraphics[width=\linewidth]{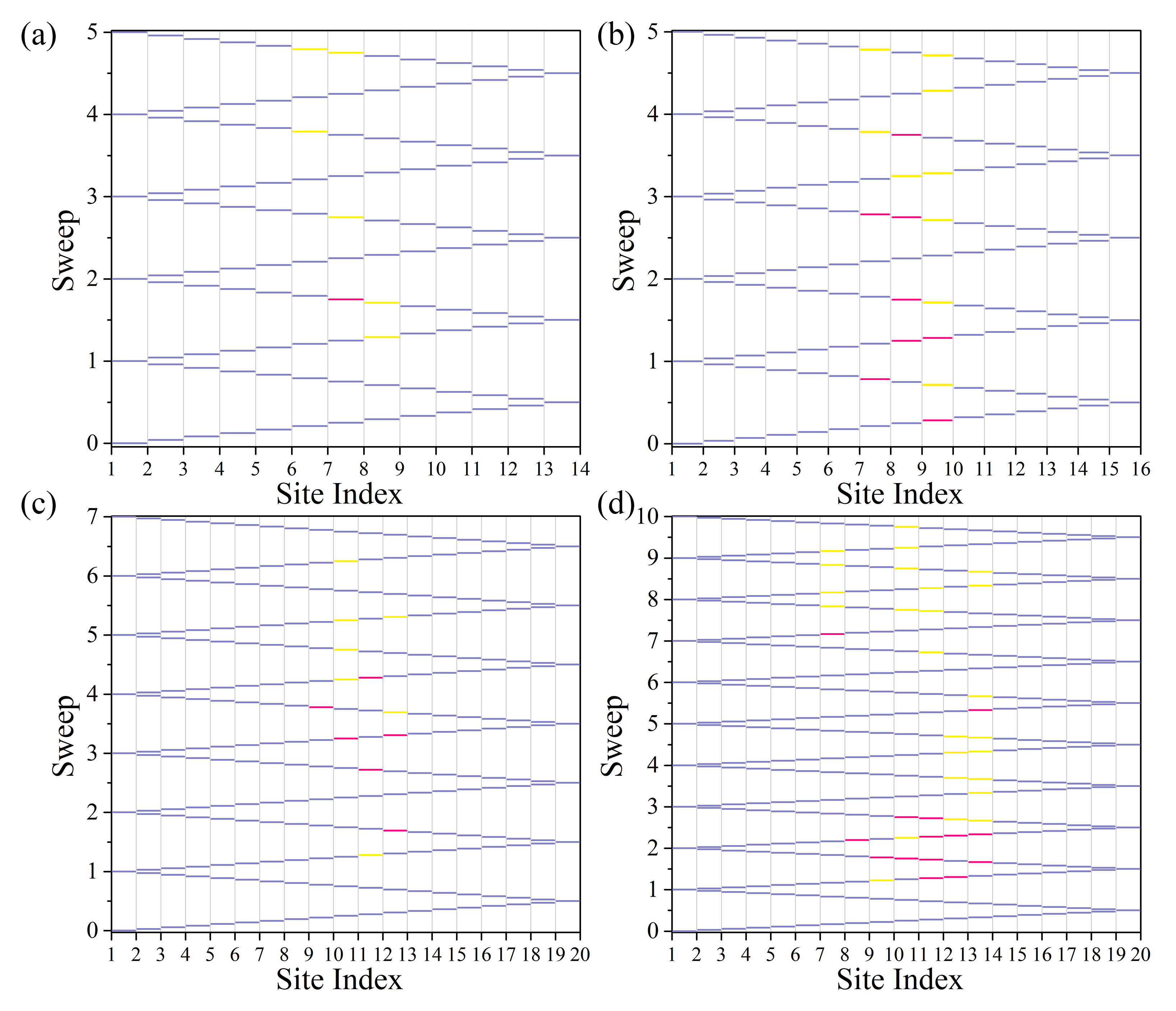}
    \caption{The optimized Clifford circuits produced in the CA-DMRG calculations for (a) H$_2$O with $\chi=40$, (b) NH$_3$ with $\chi=100$, (c) C$_2$ with $\chi=200$ and (d) N$_2$ with $\chi=100$. The yellow, red and blue lines represent those two-qubit Clifford circuits which contain a single CNOT gate, a single SWAP or iSWAP gate, only single-qubit Clifford gates, respectively.
    }
    \label{fig:clifford}
\end{figure}

It would be interesting to look at the detailed information of the optimized Clifford circuits for the different molecular systems we have studied, which are plotted in Fig.~\ref{fig:clifford}. 
The yellow lines represent those two-qubit Clifford circuits which contain a single CNOT gate, the red lines represent those two-qubit Clifford circuits which contain a single SWAP or iSWAP gate, while the rest blues represent those two-qubit Clifford circuits which contain only single-qubit Clifford gates~\cite{CorcolesSteffen2013}.
We can see that the two-qubit Clifford gates are only generated close to the center, and that the ``trivial'' SWAP gate does not contribute a dominate portion of the two-qubit gates for all the studied cases (the initial ordering of the orbitals used in this work is the default ordering provided by the PySCF package~\cite{pyscf}. If the SWAP or iSWAP gates dominate the two-qubit Clifford gates, it is likely due to that the ordering of the orbitals is not optimal, which can simply be resolved by reordering the orbitals at the beginning).
Moreover, as can be seen from Fig.~\ref{fig:clifford}(c), the SWAP and iSWAP gates are mostly located in the first $4$ sweeps for the C$_2$ molecule, while from Fig.~\ref{fig:error}(c), we can see that the energy still goes down significantly beyond $4$ sweeps. This result illustrates that the latter part of the Clifford circuit, where the two-qubit gates are mostly CNOT gates, has a clear effect on lowering the energy.

\begin{figure}
    \centering
    \includegraphics[width=\linewidth]{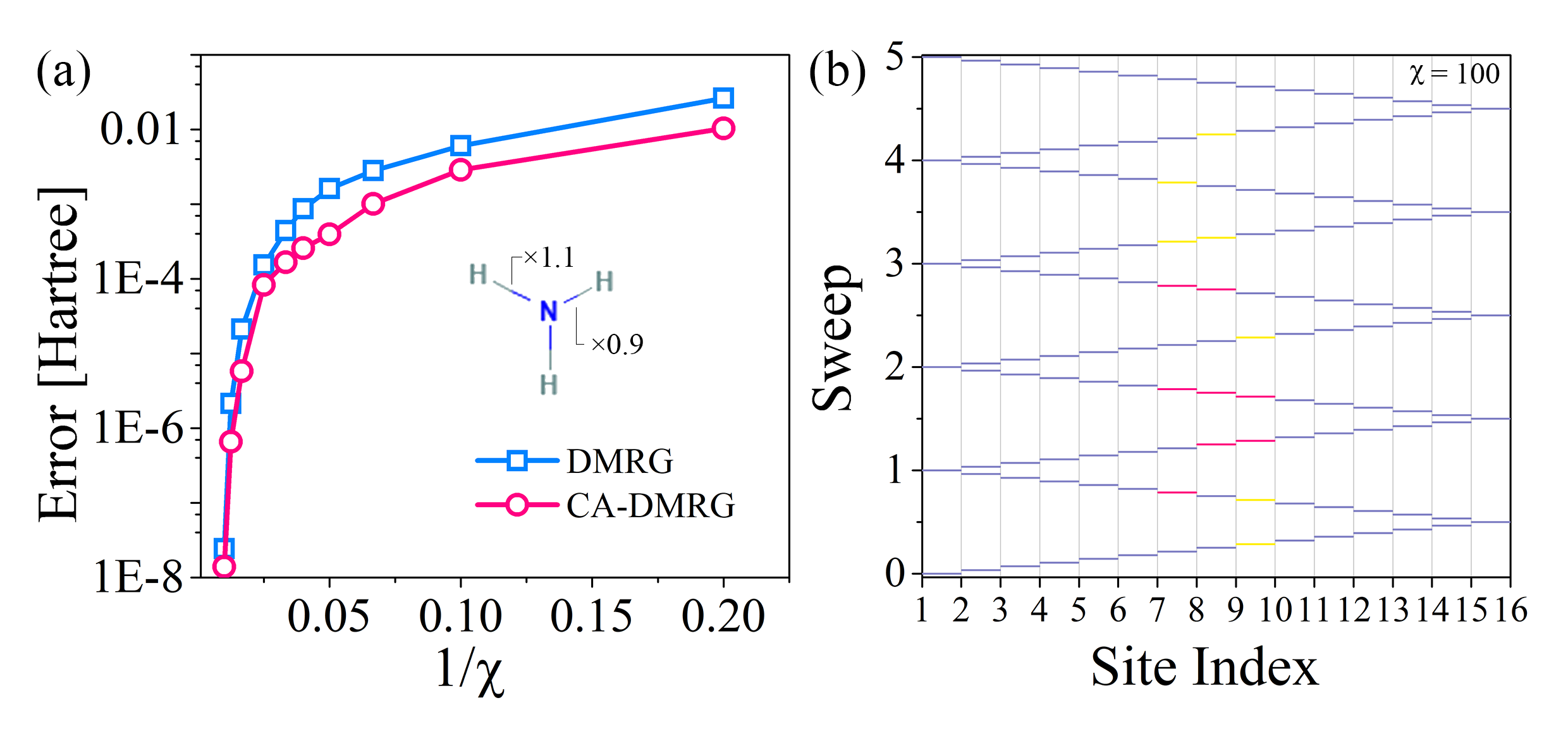}
    \caption{(a) The errors between the ground state energies and the corresponding FCI energies for the NH$_3$ with unbalanced bond lengths as a function of the inverse bond dimension. (b) The optimized Clifford circuits produced in the CA-DMRG calculation in (a) with fixed $\chi=100$.
    }
    \label{fig:energy2}
\end{figure}

To this end, we note that the molecular systems we have chosen in the previous simulations all possess certain point group symmetries.
It would also be interesting to see how the performance of CA-DMRG is if there are no such symmetries. For this purpose, we choose an ``artificial'' NH$_3$ molecule with unbalanced bond lengths to reduce the point group symmetry, and benchmark the accuracy between CA-DMRG and DMRG. The results are plotted in Fig.~\ref{fig:energy2}, where in panel (a) we plot the errors of these two methods against FCI results as a function of the inverse bond dimension, and in panel (b) we plot the optimized Clifford circuit in CA-DMRG.
Comparing Fig.~\ref{fig:energy2}(a) with Fig.~\ref{fig:energy}(b), we see that the improvement of CA-DMRG is more significant in the balanced case. 
While from Fig.~\ref{fig:energy2}(b) and Fig.~\ref{fig:clifford}(b), the pattern of the optimized Clifford circuits does not change significantly. 
Moreover, if we compare Fig.~\ref{fig:energy}(c,d) with Fig.~\ref{fig:energy}(a,b), we can also see that the improvement of CA-DMRG over DMRG is more significant for C$_2$ and N$_2$, which have higher point group symmetries. 
Therefore, our numerical results indicate that CA-DMRG is more effective for cases with higher point group symmetry.
In fact, it has been pointed out that CAMPS could better identify and encode global symmetries~\cite{MishmashMezzacapo2023}, which is in line with the observations in this work.



\section{Conclusion and discussions}
In summary, we have proposed an efficient scheme which is used in the Clifford augmented density matrix renormalization group algorithm for the purpose of dealing with \textit{ab initio} quantum chemistry Hamiltonians, such that we can largely reuse the MPO representation of the Hamiltonian during the sweeps. We apply our adapted CA-DMRG algorithm to study several molecular systems, as well as to calculate the potential energy curve of the N$_2$ molecular system. Our results show that for \textit{ab initio} quantum chemistry calculations, CA-DMRG could obtain higher accuracy than DMRG while the computational cost still remains comparable to DMRG, on par with the observations in previous applications of CA-DMRG for quantum lattice models. 

We stress that in our current implementation of CA-DMRG and DMRG, we have not used parallelization or the state-of-the-art computing hardware such as GPU~\cite{MenczerLegeza2024,MenczerFrank2025}, as such we only aim at a proof of principle demonstration of the effectiveness and accuracy of CA-DMRG for small-scale molecular systems. For DMRG, it is a standard practice to explore quantum symmetries and the sparsity of the quantum chemistry MPO for between efficiency~\cite{ChanWhite2016}. While for CA-DMRG, it is unclear whether quantum symmetries can be explored or not (the sparsity of MPO is unlikely to be explored in our approach as it is completely destroyed during canonicalization). So a fair and large-scale comparison between these two approaches still remains open.

The data related to this paper is archived at~\cite{DataRepo}.

\textit{Note added.} During our preparation of this work, we notice a recent paper which further generalizes DMRG to work with an ansatz that combines Clifford circuit, the match gates and MPS, and studies the one-dimensional hydrogen chain~\cite{HuangQin2025}.

\acknowledgements 
This work is supported by National Natural Science Foundation of China [T2222026], the Strategic Priority Research Program of the Chinese Academy of Sciences [XDB0450101].
C. G. acknowledges support from the Natural Science Foundation of Henan (Grant No. 242300421049).





\begin{thebibliography}{53}%
\makeatletter
\providecommand \@ifxundefined [1]{%
 \@ifx{#1\undefined}
}%
\providecommand \@ifnum [1]{%
 \ifnum #1\expandafter \@firstoftwo
 \else \expandafter \@secondoftwo
 \fi
}%
\providecommand \@ifx [1]{%
 \ifx #1\expandafter \@firstoftwo
 \else \expandafter \@secondoftwo
 \fi
}%
\providecommand \natexlab [1]{#1}%
\providecommand \enquote  [1]{``#1''}%
\providecommand \bibnamefont  [1]{#1}%
\providecommand \bibfnamefont [1]{#1}%
\providecommand \citenamefont [1]{#1}%
\providecommand \href@noop [0]{\@secondoftwo}%
\providecommand \href [0]{\begingroup \@sanitize@url \@href}%
\providecommand \@href[1]{\@@startlink{#1}\@@href}%
\providecommand \@@href[1]{\endgroup#1\@@endlink}%
\providecommand \@sanitize@url [0]{\catcode `\\12\catcode `\$12\catcode
  `\&12\catcode `\#12\catcode `\^12\catcode `\_12\catcode `\%12\relax}%
\providecommand \@@startlink[1]{}%
\providecommand \@@endlink[0]{}%
\providecommand \url  [0]{\begingroup\@sanitize@url \@url }%
\providecommand \@url [1]{\endgroup\@href {#1}{\urlprefix }}%
\providecommand \urlprefix  [0]{URL }%
\providecommand \Eprint [0]{\href }%
\providecommand \doibase [0]{https://doi.org/}%
\providecommand \selectlanguage [0]{\@gobble}%
\providecommand \bibinfo  [0]{\@secondoftwo}%
\providecommand \bibfield  [0]{\@secondoftwo}%
\providecommand \translation [1]{[#1]}%
\providecommand \BibitemOpen [0]{}%
\providecommand \bibitemStop [0]{}%
\providecommand \bibitemNoStop [0]{.\EOS\space}%
\providecommand \EOS [0]{\spacefactor3000\relax}%
\providecommand \BibitemShut  [1]{\csname bibitem#1\endcsname}%
\let\auto@bib@innerbib\@empty
\bibitem [{\citenamefont {Vogiatzis}\ \emph {et~al.}(2017)\citenamefont
  {Vogiatzis}, \citenamefont {Ma}, \citenamefont {Olsen}, \citenamefont
  {Gagliardi},\ and\ \citenamefont {de~Jong}}]{VogiatzisJong2017}%
  \BibitemOpen
  \bibfield  {author} {\bibinfo {author} {\bibfnamefont {K.~D.}\ \bibnamefont
  {Vogiatzis}}, \bibinfo {author} {\bibfnamefont {D.}~\bibnamefont {Ma}},
  \bibinfo {author} {\bibfnamefont {J.}~\bibnamefont {Olsen}}, \bibinfo
  {author} {\bibfnamefont {L.}~\bibnamefont {Gagliardi}},\ and\ \bibinfo
  {author} {\bibfnamefont {W.~A.}\ \bibnamefont {de~Jong}},\ }\bibfield
  {title} {\bibinfo {title} {Pushing configuration-interaction to the limit:
  Towards massively parallel mcscf calculations},\ }\href
  {https://doi.org/10.1063/1.4989858} {\bibfield  {journal} {\bibinfo
  {journal} {The Journal of Chemical Physics}\ }\textbf {\bibinfo {volume}
  {147}},\ \bibinfo {pages} {184111} (\bibinfo {year} {2017})},\ \Eprint
  {https://arxiv.org/abs/https://doi.org/10.1063/1.4989858}
  {https://doi.org/10.1063/1.4989858} \BibitemShut {NoStop}%
\bibitem [{\citenamefont {Gao}\ \emph {et~al.}(2024)\citenamefont {Gao},
  \citenamefont {Imamura}, \citenamefont {Kasagi},\ and\ \citenamefont
  {Yoshida}}]{GaoYoshida2024}%
  \BibitemOpen
  \bibfield  {author} {\bibinfo {author} {\bibfnamefont {H.}~\bibnamefont
  {Gao}}, \bibinfo {author} {\bibfnamefont {S.}~\bibnamefont {Imamura}},
  \bibinfo {author} {\bibfnamefont {A.}~\bibnamefont {Kasagi}},\ and\ \bibinfo
  {author} {\bibfnamefont {E.}~\bibnamefont {Yoshida}},\ }\bibfield  {title}
  {\bibinfo {title} {Distributed implementation of full configuration
  interaction for one trillion determinants},\ }\href
  {https://doi.org/10.1021/acs.jctc.3c01190} {\bibfield  {journal} {\bibinfo
  {journal} {Journal of Chemical Theory and Computation}\ }\textbf {\bibinfo
  {volume} {20}},\ \bibinfo {pages} {1185} (\bibinfo {year}
  {2024})}\BibitemShut {NoStop}%
\bibitem [{\citenamefont {Hartree}(1928)}]{Hartree1928}%
  \BibitemOpen
  \bibfield  {author} {\bibinfo {author} {\bibfnamefont {D.~R.}\ \bibnamefont
  {Hartree}},\ }\bibfield  {title} {\bibinfo {title} {The wave mechanics of an
  atom with a non-coulomb central field. part i. theory and methods},\ }\href
  {https://doi.org/10.1017/S0305004100011919} {\bibfield  {journal} {\bibinfo
  {journal} {Mathematical Proceedings of the Cambridge Philosophical Society}\
  }\textbf {\bibinfo {volume} {24}},\ \bibinfo {pages} {89–110} (\bibinfo
  {year} {1928})}\BibitemShut {NoStop}%
\bibitem [{\citenamefont {Fock}(1930)}]{Fock1930}%
  \BibitemOpen
  \bibfield  {author} {\bibinfo {author} {\bibfnamefont {V.}~\bibnamefont
  {Fock}},\ }\bibfield  {title} {\bibinfo {title} {N{\"a}herungsmethode zur
  l{\"o}sung des quantenmechanischen mehrk{\"o}rperproblems},\ }\href@noop {}
  {\bibfield  {journal} {\bibinfo  {journal} {Zeitschrift f{\"u}r Physik}\
  }\textbf {\bibinfo {volume} {61}},\ \bibinfo {pages} {126} (\bibinfo {year}
  {1930})}\BibitemShut {NoStop}%
\bibitem [{\citenamefont {Kohn}\ and\ \citenamefont
  {Sham}(1965)}]{KohnSham1965}%
  \BibitemOpen
  \bibfield  {author} {\bibinfo {author} {\bibfnamefont {W.}~\bibnamefont
  {Kohn}}\ and\ \bibinfo {author} {\bibfnamefont {L.~J.}\ \bibnamefont
  {Sham}},\ }\bibfield  {title} {\bibinfo {title} {Self-consistent equations
  including exchange and correlation effects},\ }\href
  {https://doi.org/10.1103/PhysRev.140.A1133} {\bibfield  {journal} {\bibinfo
  {journal} {Phys. Rev.}\ }\textbf {\bibinfo {volume} {140}},\ \bibinfo {pages}
  {A1133} (\bibinfo {year} {1965})}\BibitemShut {NoStop}%
\bibitem [{\citenamefont {Holmes}\ \emph {et~al.}(2016)\citenamefont {Holmes},
  \citenamefont {Tubman},\ and\ \citenamefont {Umrigar}}]{HolmesUmrigar2016}%
  \BibitemOpen
  \bibfield  {author} {\bibinfo {author} {\bibfnamefont {A.~A.}\ \bibnamefont
  {Holmes}}, \bibinfo {author} {\bibfnamefont {N.~M.}\ \bibnamefont {Tubman}},\
  and\ \bibinfo {author} {\bibfnamefont {C.~J.}\ \bibnamefont {Umrigar}},\
  }\bibfield  {title} {\bibinfo {title} {Heat-bath configuration interaction:
  An efficient selected configuration interaction algorithm inspired by
  heat-bath sampling},\ }\href {https://doi.org/10.1021/acs.jctc.6b00407}
  {\bibfield  {journal} {\bibinfo  {journal} {Journal of Chemical Theory and
  Computation}\ }\textbf {\bibinfo {volume} {12}},\ \bibinfo {pages}
  {3674–3680} (\bibinfo {year} {2016})}\BibitemShut {NoStop}%
\bibitem [{\citenamefont {Schriber}\ and\ \citenamefont
  {Evangelista}(2017)}]{SchriberEvangelista2017}%
  \BibitemOpen
  \bibfield  {author} {\bibinfo {author} {\bibfnamefont {J.~B.}\ \bibnamefont
  {Schriber}}\ and\ \bibinfo {author} {\bibfnamefont {F.~A.}\ \bibnamefont
  {Evangelista}},\ }\bibfield  {title} {\bibinfo {title} {Adaptive
  configuration interaction for computing challenging electronic excited states
  with tunable accuracy},\ }\href@noop {} {\bibfield  {journal} {\bibinfo
  {journal} {Journal of chemical theory and computation}\ }\textbf {\bibinfo
  {volume} {13}},\ \bibinfo {pages} {5354} (\bibinfo {year}
  {2017})}\BibitemShut {NoStop}%
\bibitem [{\citenamefont {Purvis}\ and\ \citenamefont
  {Bartlett}(1982)}]{CCSD-Purvis1982AFC}%
  \BibitemOpen
  \bibfield  {author} {\bibinfo {author} {\bibfnamefont {G.~D.}\ \bibnamefont
  {Purvis}}\ and\ \bibinfo {author} {\bibfnamefont {R.~J.}\ \bibnamefont
  {Bartlett}},\ }\bibfield  {title} {\bibinfo {title} {A full coupled‐cluster
  singles and doubles model: The inclusion of disconnected triples},\
  }\href@noop {} {\bibfield  {journal} {\bibinfo  {journal} {Journal of
  Chemical Physics}\ }\textbf {\bibinfo {volume} {76}},\ \bibinfo {pages}
  {1910} (\bibinfo {year} {1982})}\BibitemShut {NoStop}%
\bibitem [{\citenamefont {Foulkes}\ \emph {et~al.}(2001)\citenamefont
  {Foulkes}, \citenamefont {Mitas}, \citenamefont {Needs},\ and\ \citenamefont
  {Rajagopal}}]{FoulkesRajagopal2001}%
  \BibitemOpen
  \bibfield  {author} {\bibinfo {author} {\bibfnamefont {W.~M.~C.}\
  \bibnamefont {Foulkes}}, \bibinfo {author} {\bibfnamefont {L.}~\bibnamefont
  {Mitas}}, \bibinfo {author} {\bibfnamefont {R.~J.}\ \bibnamefont {Needs}},\
  and\ \bibinfo {author} {\bibfnamefont {G.}~\bibnamefont {Rajagopal}},\
  }\bibfield  {title} {\bibinfo {title} {Quantum monte carlo simulations of
  solids},\ }\href {https://doi.org/10.1103/RevModPhys.73.33} {\bibfield
  {journal} {\bibinfo  {journal} {Rev. Mod. Phys.}\ }\textbf {\bibinfo {volume}
  {73}},\ \bibinfo {pages} {33} (\bibinfo {year} {2001})}\BibitemShut {NoStop}%
\bibitem [{\citenamefont {White}(1992)}]{White1992}%
  \BibitemOpen
  \bibfield  {author} {\bibinfo {author} {\bibfnamefont {S.~R.}\ \bibnamefont
  {White}},\ }\bibfield  {title} {\bibinfo {title} {Density matrix formulation
  for quantum renormalization groups},\ }\href
  {https://doi.org/10.1103/PhysRevLett.69.2863} {\bibfield  {journal} {\bibinfo
   {journal} {Phys. Rev. Lett.}\ }\textbf {\bibinfo {volume} {69}},\ \bibinfo
  {pages} {2863} (\bibinfo {year} {1992})}\BibitemShut {NoStop}%
\bibitem [{\citenamefont {White}(1993)}]{White1993}%
  \BibitemOpen
  \bibfield  {author} {\bibinfo {author} {\bibfnamefont {S.~R.}\ \bibnamefont
  {White}},\ }\bibfield  {title} {\bibinfo {title} {Density-matrix algorithms
  for quantum renormalization groups},\ }\href
  {https://doi.org/10.1103/PhysRevB.48.10345} {\bibfield  {journal} {\bibinfo
  {journal} {Phys. Rev. B}\ }\textbf {\bibinfo {volume} {48}},\ \bibinfo
  {pages} {10345} (\bibinfo {year} {1993})}\BibitemShut {NoStop}%
\bibitem [{\citenamefont {Carleo}\ and\ \citenamefont
  {Troyer}(2017)}]{CarleoTroyer2017}%
  \BibitemOpen
  \bibfield  {author} {\bibinfo {author} {\bibfnamefont {G.}~\bibnamefont
  {Carleo}}\ and\ \bibinfo {author} {\bibfnamefont {M.}~\bibnamefont
  {Troyer}},\ }\bibfield  {title} {\bibinfo {title} {Solving the quantum
  many-body problem with artificial neural networks},\ }\href
  {https://doi.org/10.1126/science.aag2302} {\bibfield  {journal} {\bibinfo
  {journal} {Science}\ }\textbf {\bibinfo {volume} {355}},\ \bibinfo {pages}
  {602} (\bibinfo {year} {2017})},\ \Eprint
  {https://arxiv.org/abs/https://www.science.org/doi/pdf/10.1126/science.aag2302}
  {https://www.science.org/doi/pdf/10.1126/science.aag2302} \BibitemShut
  {NoStop}%
\bibitem [{\citenamefont {Choo}\ \emph {et~al.}(2020)\citenamefont {Choo},
  \citenamefont {Mezzacapo},\ and\ \citenamefont {Carleo}}]{ChooCarleo2020}%
  \BibitemOpen
  \bibfield  {author} {\bibinfo {author} {\bibfnamefont {K.}~\bibnamefont
  {Choo}}, \bibinfo {author} {\bibfnamefont {A.}~\bibnamefont {Mezzacapo}},\
  and\ \bibinfo {author} {\bibfnamefont {G.}~\bibnamefont {Carleo}},\
  }\bibfield  {title} {\bibinfo {title} {Fermionic neural-network states for
  ab-initio electronic structure},\ }\href
  {https://doi.org/10.1038/s41467-020-15724-9} {\bibfield  {journal} {\bibinfo
  {journal} {Nature communications}\ }\textbf {\bibinfo {volume} {11}},\
  \bibinfo {pages} {2368} (\bibinfo {year} {2020})}\BibitemShut {NoStop}%
\bibitem [{\citenamefont {Hermann}\ \emph {et~al.}(2020)\citenamefont
  {Hermann}, \citenamefont {Sch{\"a}tzle},\ and\ \citenamefont
  {No{\'e}}}]{HermannNoe2020}%
  \BibitemOpen
  \bibfield  {author} {\bibinfo {author} {\bibfnamefont {J.}~\bibnamefont
  {Hermann}}, \bibinfo {author} {\bibfnamefont {Z.}~\bibnamefont
  {Sch{\"a}tzle}},\ and\ \bibinfo {author} {\bibfnamefont {F.}~\bibnamefont
  {No{\'e}}},\ }\bibfield  {title} {\bibinfo {title} {Deep-neural-network
  solution of the electronic schr{\"o}dinger equation},\ }\href
  {https://doi.org/10.1038/s41557-020-0544-y} {\bibfield  {journal} {\bibinfo
  {journal} {Nature Chemistry}\ }\textbf {\bibinfo {volume} {12}},\ \bibinfo
  {pages} {891} (\bibinfo {year} {2020})}\BibitemShut {NoStop}%
\bibitem [{\citenamefont {Pfau}\ \emph {et~al.}(2020)\citenamefont {Pfau},
  \citenamefont {Spencer}, \citenamefont {Matthews},\ and\ \citenamefont
  {Foulkes}}]{PfauFoulkes2020}%
  \BibitemOpen
  \bibfield  {author} {\bibinfo {author} {\bibfnamefont {D.}~\bibnamefont
  {Pfau}}, \bibinfo {author} {\bibfnamefont {J.~S.}\ \bibnamefont {Spencer}},
  \bibinfo {author} {\bibfnamefont {A.~G. D.~G.}\ \bibnamefont {Matthews}},\
  and\ \bibinfo {author} {\bibfnamefont {W.~M.~C.}\ \bibnamefont {Foulkes}},\
  }\bibfield  {title} {\bibinfo {title} {Ab initio solution of the
  many-electron schr\"odinger equation with deep neural networks},\ }\href
  {https://doi.org/10.1103/PhysRevResearch.2.033429} {\bibfield  {journal}
  {\bibinfo  {journal} {Phys. Rev. Res.}\ }\textbf {\bibinfo {volume} {2}},\
  \bibinfo {pages} {033429} (\bibinfo {year} {2020})}\BibitemShut {NoStop}%
\bibitem [{\citenamefont {Barrett}\ \emph {et~al.}(2022)\citenamefont
  {Barrett}, \citenamefont {Malyshev},\ and\ \citenamefont
  {Lvovsky}}]{BarrettLvovsky2022}%
  \BibitemOpen
  \bibfield  {author} {\bibinfo {author} {\bibfnamefont {T.~D.}\ \bibnamefont
  {Barrett}}, \bibinfo {author} {\bibfnamefont {A.}~\bibnamefont {Malyshev}},\
  and\ \bibinfo {author} {\bibfnamefont {A.}~\bibnamefont {Lvovsky}},\
  }\bibfield  {title} {\bibinfo {title} {Autoregressive neural-network
  wavefunctions for ab initio quantum chemistry},\ }\href
  {https://doi.org/10.1038/s42256-022-00461-z} {\bibfield  {journal} {\bibinfo
  {journal} {Nature Machine Intelligence}\ }\textbf {\bibinfo {volume} {4}},\
  \bibinfo {pages} {351} (\bibinfo {year} {2022})}\BibitemShut {NoStop}%
\bibitem [{\citenamefont {Wu}\ \emph {et~al.}(2023{\natexlab{a}})\citenamefont
  {Wu}, \citenamefont {Guo}, \citenamefont {Fan}, \citenamefont {Zhou},\ and\
  \citenamefont {Shang}}]{WuShang2023}%
  \BibitemOpen
  \bibfield  {author} {\bibinfo {author} {\bibfnamefont {Y.}~\bibnamefont
  {Wu}}, \bibinfo {author} {\bibfnamefont {C.}~\bibnamefont {Guo}}, \bibinfo
  {author} {\bibfnamefont {Y.}~\bibnamefont {Fan}}, \bibinfo {author}
  {\bibfnamefont {P.}~\bibnamefont {Zhou}},\ and\ \bibinfo {author}
  {\bibfnamefont {H.}~\bibnamefont {Shang}},\ }\bibfield  {title} {\bibinfo
  {title} {Nnqs-transformer: an efficient and scalable neural network quantum
  states approach for ab initio quantum chemistry},\ }in\ \href
  {https://doi.org/10.1145/3581784.3607061} {\emph {\bibinfo {booktitle}
  {Proceedings of the International Conference for High Performance Computing,
  Networking, Storage and Analysis}}},\ \bibinfo {series and number} {SC '23}\
  (\bibinfo  {publisher} {Association for Computing Machinery},\ \bibinfo
  {address} {New York, NY, USA},\ \bibinfo {year} {2023})\BibitemShut {NoStop}%
\bibitem [{\citenamefont {von Glehn}\ \emph {et~al.}(2022)\citenamefont {von
  Glehn}, \citenamefont {Spencer},\ and\ \citenamefont {Pfau}}]{VonPfau2022}%
  \BibitemOpen
  \bibfield  {author} {\bibinfo {author} {\bibfnamefont {I.}~\bibnamefont {von
  Glehn}}, \bibinfo {author} {\bibfnamefont {J.~S.}\ \bibnamefont {Spencer}},\
  and\ \bibinfo {author} {\bibfnamefont {D.}~\bibnamefont {Pfau}},\ }\bibfield
  {title} {\bibinfo {title} {A self-attention ansatz for ab-initio quantum
  chemistry},\ }\href {https://arxiv.org/abs/2211.13672} {\bibfield  {journal}
  {\bibinfo  {journal} {arXiv:2211.13672}\ } (\bibinfo {year}
  {2022})}\BibitemShut {NoStop}%
\bibitem [{\citenamefont {Xiang}(1996)}]{Xiang1996}%
  \BibitemOpen
  \bibfield  {author} {\bibinfo {author} {\bibfnamefont {T.}~\bibnamefont
  {Xiang}},\ }\bibfield  {title} {\bibinfo {title} {Density-matrix
  renormalization-group method in momentum space},\ }\href
  {https://doi.org/10.1103/PhysRevB.53.R10445} {\bibfield  {journal} {\bibinfo
  {journal} {Phys. Rev. B}\ }\textbf {\bibinfo {volume} {53}},\ \bibinfo
  {pages} {R10445} (\bibinfo {year} {1996})}\BibitemShut {NoStop}%
\bibitem [{\citenamefont {White}\ and\ \citenamefont
  {Martin}(1999)}]{WhiteMartin1999}%
  \BibitemOpen
  \bibfield  {author} {\bibinfo {author} {\bibfnamefont {S.~R.}\ \bibnamefont
  {White}}\ and\ \bibinfo {author} {\bibfnamefont {R.~L.}\ \bibnamefont
  {Martin}},\ }\bibfield  {title} {\bibinfo {title} {Ab initio quantum
  chemistry using the density matrix renormalization group},\ }\href
  {https://doi.org/10.1063/1.478295} {\bibfield  {journal} {\bibinfo  {journal}
  {The Journal of Chemical Physics}\ }\textbf {\bibinfo {volume} {110}},\
  \bibinfo {pages} {4127} (\bibinfo {year} {1999})},\ \Eprint
  {https://arxiv.org/abs/https://pubs.aip.org/aip/jcp/article-pdf/110/9/4127/19023329/4127\_1\_online.pdf}
  {https://pubs.aip.org/aip/jcp/article-pdf/110/9/4127/19023329/4127\_1\_online.pdf}
  \BibitemShut {NoStop}%
\bibitem [{\citenamefont {Marti}\ \emph {et~al.}(2008)\citenamefont {Marti},
  \citenamefont {Ondík}, \citenamefont {Moritz},\ and\ \citenamefont
  {Reiher}}]{MartiReiher2008}%
  \BibitemOpen
  \bibfield  {author} {\bibinfo {author} {\bibfnamefont {K.~H.}\ \bibnamefont
  {Marti}}, \bibinfo {author} {\bibfnamefont {I.~M.}\ \bibnamefont {Ondík}},
  \bibinfo {author} {\bibfnamefont {G.}~\bibnamefont {Moritz}},\ and\ \bibinfo
  {author} {\bibfnamefont {M.}~\bibnamefont {Reiher}},\ }\bibfield  {title}
  {\bibinfo {title} {Density matrix renormalization group calculations on
  relative energies of transition metal complexes and clusters},\ }\href
  {https://doi.org/10.1063/1.2805383} {\bibfield  {journal} {\bibinfo
  {journal} {The Journal of Chemical Physics}\ }\textbf {\bibinfo {volume}
  {128}},\ \bibinfo {pages} {014104} (\bibinfo {year} {2008})},\ \Eprint
  {https://arxiv.org/abs/https://pubs.aip.org/aip/jcp/article-pdf/doi/10.1063/1.2805383/15408073/014104\_1\_online.pdf}
  {https://pubs.aip.org/aip/jcp/article-pdf/doi/10.1063/1.2805383/15408073/014104\_1\_online.pdf}
  \BibitemShut {NoStop}%
\bibitem [{\citenamefont {Chan}\ and\ \citenamefont
  {Sharma}(2011)}]{ChanSharma2011}%
  \BibitemOpen
  \bibfield  {author} {\bibinfo {author} {\bibfnamefont {G.~K.-L.}\
  \bibnamefont {Chan}}\ and\ \bibinfo {author} {\bibfnamefont {S.}~\bibnamefont
  {Sharma}},\ }\bibfield  {title} {\bibinfo {title} {The density matrix
  renormalization group in quantum chemistry},\ }\href
  {https://doi.org/https://doi.org/10.1146/annurev-physchem-032210-103338}
  {\bibfield  {journal} {\bibinfo  {journal} {Annual Review of Physical
  Chemistry}\ }\textbf {\bibinfo {volume} {62}},\ \bibinfo {pages} {465}
  (\bibinfo {year} {2011})}\BibitemShut {NoStop}%
\bibitem [{\citenamefont {Kurashige}\ \emph {et~al.}(2013)\citenamefont
  {Kurashige}, \citenamefont {Chan},\ and\ \citenamefont
  {Yanai}}]{KurashigeYanai2013}%
  \BibitemOpen
  \bibfield  {author} {\bibinfo {author} {\bibfnamefont {Y.}~\bibnamefont
  {Kurashige}}, \bibinfo {author} {\bibfnamefont {G.~K.-L.}\ \bibnamefont
  {Chan}},\ and\ \bibinfo {author} {\bibfnamefont {T.}~\bibnamefont {Yanai}},\
  }\bibfield  {title} {\bibinfo {title} {Entangled quantum electronic
  wavefunctions of the mn4cao5 cluster in photosystem ii},\ }\href
  {https://doi.org/10.1038/nchem.1677} {\bibfield  {journal} {\bibinfo
  {journal} {Nature chemistry}\ }\textbf {\bibinfo {volume} {5}},\ \bibinfo
  {pages} {660} (\bibinfo {year} {2013})}\BibitemShut {NoStop}%
\bibitem [{\citenamefont {Wouters}\ \emph {et~al.}(2014)\citenamefont
  {Wouters}, \citenamefont {Poelmans}, \citenamefont {Ayers},\ and\
  \citenamefont {{Van Neck}}}]{WoutersNeck2014}%
  \BibitemOpen
  \bibfield  {author} {\bibinfo {author} {\bibfnamefont {S.}~\bibnamefont
  {Wouters}}, \bibinfo {author} {\bibfnamefont {W.}~\bibnamefont {Poelmans}},
  \bibinfo {author} {\bibfnamefont {P.~W.}\ \bibnamefont {Ayers}},\ and\
  \bibinfo {author} {\bibfnamefont {D.}~\bibnamefont {{Van Neck}}},\ }\bibfield
   {title} {\bibinfo {title} {Chemps2: A free open-source spin-adapted
  implementation of the density matrix renormalization group for ab initio
  quantum chemistry},\ }\href
  {https://doi.org/https://doi.org/10.1016/j.cpc.2014.01.019} {\bibfield
  {journal} {\bibinfo  {journal} {Computer Physics Communications}\ }\textbf
  {\bibinfo {volume} {185}},\ \bibinfo {pages} {1501} (\bibinfo {year}
  {2014})}\BibitemShut {NoStop}%
\bibitem [{\citenamefont {Li}\ \emph {et~al.}(2019)\citenamefont {Li},
  \citenamefont {Guo}, \citenamefont {Sun},\ and\ \citenamefont
  {Chan}}]{LiChan2019}%
  \BibitemOpen
  \bibfield  {author} {\bibinfo {author} {\bibfnamefont {Z.}~\bibnamefont
  {Li}}, \bibinfo {author} {\bibfnamefont {S.}~\bibnamefont {Guo}}, \bibinfo
  {author} {\bibfnamefont {Q.}~\bibnamefont {Sun}},\ and\ \bibinfo {author}
  {\bibfnamefont {G.~K.-L.}\ \bibnamefont {Chan}},\ }\bibfield  {title}
  {\bibinfo {title} {Electronic landscape of the p-cluster of nitrogenase as
  revealed through many-electron quantum wavefunction simulations},\ }\href
  {https://doi.org/10.1038/s41557-019-0337-3} {\bibfield  {journal} {\bibinfo
  {journal} {Nature chemistry}\ }\textbf {\bibinfo {volume} {11}},\ \bibinfo
  {pages} {1026} (\bibinfo {year} {2019})}\BibitemShut {NoStop}%
\bibitem [{\citenamefont {Larsson}\ \emph {et~al.}(2022)\citenamefont
  {Larsson}, \citenamefont {Zhai}, \citenamefont {Umrigar},\ and\ \citenamefont
  {Chan}}]{LarssonChan2022}%
  \BibitemOpen
  \bibfield  {author} {\bibinfo {author} {\bibfnamefont {H.~R.}\ \bibnamefont
  {Larsson}}, \bibinfo {author} {\bibfnamefont {H.}~\bibnamefont {Zhai}},
  \bibinfo {author} {\bibfnamefont {C.~J.}\ \bibnamefont {Umrigar}},\ and\
  \bibinfo {author} {\bibfnamefont {G.~K.-L.}\ \bibnamefont {Chan}},\
  }\bibfield  {title} {\bibinfo {title} {The chromium dimer: closing a chapter
  of quantum chemistry},\ }\href {https://doi.org/10.1021/jacs.2c06357}
  {\bibfield  {journal} {\bibinfo  {journal} {Journal of the American Chemical
  Society}\ }\textbf {\bibinfo {volume} {144}},\ \bibinfo {pages} {15932}
  (\bibinfo {year} {2022})}\BibitemShut {NoStop}%
\bibitem [{\citenamefont {Xiang}\ \emph {et~al.}(2024)\citenamefont {Xiang},
  \citenamefont {Jia}, \citenamefont {Fang},\ and\ \citenamefont
  {Li}}]{XiangLi2023}%
  \BibitemOpen
  \bibfield  {author} {\bibinfo {author} {\bibfnamefont {C.}~\bibnamefont
  {Xiang}}, \bibinfo {author} {\bibfnamefont {W.}~\bibnamefont {Jia}}, \bibinfo
  {author} {\bibfnamefont {W.-H.}\ \bibnamefont {Fang}},\ and\ \bibinfo
  {author} {\bibfnamefont {Z.}~\bibnamefont {Li}},\ }\bibfield  {title}
  {\bibinfo {title} {Distributed multi-gpu ab initio density matrix
  renormalization group algorithm with applications to the p-cluster of
  nitrogenase},\ }\href {https://doi.org/10.1021/acs.jctc.3c01228} {\bibfield
  {journal} {\bibinfo  {journal} {Journal of Chemical Theory and Computation}\
  }\textbf {\bibinfo {volume} {20}},\ \bibinfo {pages} {775} (\bibinfo {year}
  {2024})}\BibitemShut {NoStop}%
\bibitem [{\citenamefont {Guo}(2022)}]{Guo2022b}%
  \BibitemOpen
  \bibfield  {author} {\bibinfo {author} {\bibfnamefont {C.}~\bibnamefont
  {Guo}},\ }\bibfield  {title} {\bibinfo {title} {Density matrix
  renormalization group algorithm for mixed quantum states},\ }\href
  {https://doi.org/10.1103/PhysRevB.105.195152} {\bibfield  {journal} {\bibinfo
   {journal} {Phys. Rev. B}\ }\textbf {\bibinfo {volume} {105}},\ \bibinfo
  {pages} {195152} (\bibinfo {year} {2022})}\BibitemShut {NoStop}%
\bibitem [{\citenamefont {Deng}\ \emph {et~al.}(2017)\citenamefont {Deng},
  \citenamefont {Li},\ and\ \citenamefont {Das~Sarma}}]{DengSarma2017}%
  \BibitemOpen
  \bibfield  {author} {\bibinfo {author} {\bibfnamefont {D.-L.}\ \bibnamefont
  {Deng}}, \bibinfo {author} {\bibfnamefont {X.}~\bibnamefont {Li}},\ and\
  \bibinfo {author} {\bibfnamefont {S.}~\bibnamefont {Das~Sarma}},\ }\bibfield
  {title} {\bibinfo {title} {Quantum entanglement in neural network states},\
  }\href {https://doi.org/10.1103/PhysRevX.7.021021} {\bibfield  {journal}
  {\bibinfo  {journal} {Phys. Rev. X}\ }\textbf {\bibinfo {volume} {7}},\
  \bibinfo {pages} {021021} (\bibinfo {year} {2017})}\BibitemShut {NoStop}%
\bibitem [{\citenamefont {Sharir}\ \emph {et~al.}(2022)\citenamefont {Sharir},
  \citenamefont {Shashua},\ and\ \citenamefont {Carleo}}]{SharirCarleo2022}%
  \BibitemOpen
  \bibfield  {author} {\bibinfo {author} {\bibfnamefont {O.}~\bibnamefont
  {Sharir}}, \bibinfo {author} {\bibfnamefont {A.}~\bibnamefont {Shashua}},\
  and\ \bibinfo {author} {\bibfnamefont {G.}~\bibnamefont {Carleo}},\
  }\bibfield  {title} {\bibinfo {title} {Neural tensor contractions and the
  expressive power of deep neural quantum states},\ }\href
  {https://doi.org/10.1103/PhysRevB.106.205136} {\bibfield  {journal} {\bibinfo
   {journal} {Phys. Rev. B}\ }\textbf {\bibinfo {volume} {106}},\ \bibinfo
  {pages} {205136} (\bibinfo {year} {2022})}\BibitemShut {NoStop}%
\bibitem [{\citenamefont {Wu}\ \emph {et~al.}(2023{\natexlab{b}})\citenamefont
  {Wu}, \citenamefont {Rossi}, \citenamefont {Vicentini},\ and\ \citenamefont
  {Carleo}}]{WuCarleo2023}%
  \BibitemOpen
  \bibfield  {author} {\bibinfo {author} {\bibfnamefont {D.}~\bibnamefont
  {Wu}}, \bibinfo {author} {\bibfnamefont {R.}~\bibnamefont {Rossi}}, \bibinfo
  {author} {\bibfnamefont {F.}~\bibnamefont {Vicentini}},\ and\ \bibinfo
  {author} {\bibfnamefont {G.}~\bibnamefont {Carleo}},\ }\bibfield  {title}
  {\bibinfo {title} {From tensor-network quantum states to tensorial recurrent
  neural networks},\ }\href {https://doi.org/10.1103/PhysRevResearch.5.L032001}
  {\bibfield  {journal} {\bibinfo  {journal} {Phys. Rev. Res.}\ }\textbf
  {\bibinfo {volume} {5}},\ \bibinfo {pages} {L032001} (\bibinfo {year}
  {2023}{\natexlab{b}})}\BibitemShut {NoStop}%
\bibitem [{\citenamefont {Mishmash}\ \emph {et~al.}(2023)\citenamefont
  {Mishmash}, \citenamefont {Gujarati}, \citenamefont {Motta}, \citenamefont
  {Zhai}, \citenamefont {Chan},\ and\ \citenamefont
  {Mezzacapo}}]{MishmashMezzacapo2023}%
  \BibitemOpen
  \bibfield  {author} {\bibinfo {author} {\bibfnamefont {R.~V.}\ \bibnamefont
  {Mishmash}}, \bibinfo {author} {\bibfnamefont {T.~P.}\ \bibnamefont
  {Gujarati}}, \bibinfo {author} {\bibfnamefont {M.}~\bibnamefont {Motta}},
  \bibinfo {author} {\bibfnamefont {H.}~\bibnamefont {Zhai}}, \bibinfo {author}
  {\bibfnamefont {G.~K.-L.}\ \bibnamefont {Chan}},\ and\ \bibinfo {author}
  {\bibfnamefont {A.}~\bibnamefont {Mezzacapo}},\ }\bibfield  {title} {\bibinfo
  {title} {Hierarchical clifford transformations to reduce entanglement in
  quantum chemistry wave functions},\ }\href
  {https://doi.org/10.1021/acs.jctc.3c00228} {\bibfield  {journal} {\bibinfo
  {journal} {Journal of chemical theory and computation}\ }\textbf {\bibinfo
  {volume} {19}},\ \bibinfo {pages} {3194} (\bibinfo {year}
  {2023})}\BibitemShut {NoStop}%
\bibitem [{\citenamefont {Masot-Llima}\ and\ \citenamefont
  {Garcia-Saez}(2024)}]{LlimaSaez2024}%
  \BibitemOpen
  \bibfield  {author} {\bibinfo {author} {\bibfnamefont {S.}~\bibnamefont
  {Masot-Llima}}\ and\ \bibinfo {author} {\bibfnamefont {A.}~\bibnamefont
  {Garcia-Saez}},\ }\bibfield  {title} {\bibinfo {title} {Stabilizer tensor
  networks: Universal quantum simulator on a basis of stabilizer states},\
  }\href {https://doi.org/10.1103/PhysRevLett.133.230601} {\bibfield  {journal}
  {\bibinfo  {journal} {Phys. Rev. Lett.}\ }\textbf {\bibinfo {volume} {133}},\
  \bibinfo {pages} {230601} (\bibinfo {year} {2024})}\BibitemShut {NoStop}%
\bibitem [{\citenamefont {Lami}\ \emph {et~al.}(2025)\citenamefont {Lami},
  \citenamefont {Haug},\ and\ \citenamefont {De~Nardis}}]{LamiNardis2025}%
  \BibitemOpen
  \bibfield  {author} {\bibinfo {author} {\bibfnamefont {G.}~\bibnamefont
  {Lami}}, \bibinfo {author} {\bibfnamefont {T.}~\bibnamefont {Haug}},\ and\
  \bibinfo {author} {\bibfnamefont {J.}~\bibnamefont {De~Nardis}},\ }\bibfield
  {title} {\bibinfo {title} {Quantum state designs with clifford-enhanced
  matrix product states},\ }\href {https://doi.org/10.1103/PRXQuantum.6.010345}
  {\bibfield  {journal} {\bibinfo  {journal} {PRX Quantum}\ }\textbf {\bibinfo
  {volume} {6}},\ \bibinfo {pages} {010345} (\bibinfo {year}
  {2025})}\BibitemShut {NoStop}%
\bibitem [{\citenamefont {Gottesman}(1997)}]{Gottesman1997}%
  \BibitemOpen
  \bibfield  {author} {\bibinfo {author} {\bibfnamefont {D.}~\bibnamefont
  {Gottesman}},\ }\href@noop {} {\emph {\bibinfo {title} {Stabilizer codes and
  quantum error correction}}}\ (\bibinfo  {publisher} {California Institute of
  Technology},\ \bibinfo {year} {1997})\BibitemShut {NoStop}%
\bibitem [{\citenamefont {Aaronson}\ and\ \citenamefont
  {Gottesman}(2004)}]{AaronsonGottesman2004}%
  \BibitemOpen
  \bibfield  {author} {\bibinfo {author} {\bibfnamefont {S.}~\bibnamefont
  {Aaronson}}\ and\ \bibinfo {author} {\bibfnamefont {D.}~\bibnamefont
  {Gottesman}},\ }\bibfield  {title} {\bibinfo {title} {Improved simulation of
  stabilizer circuits},\ }\href {https://doi.org/10.1103/PhysRevA.70.052328}
  {\bibfield  {journal} {\bibinfo  {journal} {Phys. Rev. A}\ }\textbf {\bibinfo
  {volume} {70}},\ \bibinfo {pages} {052328} (\bibinfo {year}
  {2004})}\BibitemShut {NoStop}%
\bibitem [{\citenamefont {Anders}\ and\ \citenamefont
  {Briegel}(2006)}]{AndersBriegel2006}%
  \BibitemOpen
  \bibfield  {author} {\bibinfo {author} {\bibfnamefont {S.}~\bibnamefont
  {Anders}}\ and\ \bibinfo {author} {\bibfnamefont {H.~J.}\ \bibnamefont
  {Briegel}},\ }\bibfield  {title} {\bibinfo {title} {Fast simulation of
  stabilizer circuits using a graph-state representation},\ }\href
  {https://doi.org/10.1103/PhysRevA.73.022334} {\bibfield  {journal} {\bibinfo
  {journal} {Phys. Rev. A}\ }\textbf {\bibinfo {volume} {73}},\ \bibinfo
  {pages} {022334} (\bibinfo {year} {2006})}\BibitemShut {NoStop}%
\bibitem [{\citenamefont {Qian}\ \emph {et~al.}(2024)\citenamefont {Qian},
  \citenamefont {Huang},\ and\ \citenamefont {Qin}}]{QianQin2024}%
  \BibitemOpen
  \bibfield  {author} {\bibinfo {author} {\bibfnamefont {X.}~\bibnamefont
  {Qian}}, \bibinfo {author} {\bibfnamefont {J.}~\bibnamefont {Huang}},\ and\
  \bibinfo {author} {\bibfnamefont {M.}~\bibnamefont {Qin}},\ }\bibfield
  {title} {\bibinfo {title} {Augmenting density matrix renormalization group
  with clifford circuits},\ }\href
  {https://doi.org/10.1103/PhysRevLett.133.190402} {\bibfield  {journal}
  {\bibinfo  {journal} {Phys. Rev. Lett.}\ }\textbf {\bibinfo {volume} {133}},\
  \bibinfo {pages} {190402} (\bibinfo {year} {2024})}\BibitemShut {NoStop}%
\bibitem [{\citenamefont {Fan}\ \emph {et~al.}(2025)\citenamefont {Fan},
  \citenamefont {Qian}, \citenamefont {Zhang}, \citenamefont {Huang},
  \citenamefont {Qin},\ and\ \citenamefont {Xiang}}]{FanXiang2025}%
  \BibitemOpen
  \bibfield  {author} {\bibinfo {author} {\bibfnamefont {C.}~\bibnamefont
  {Fan}}, \bibinfo {author} {\bibfnamefont {X.}~\bibnamefont {Qian}}, \bibinfo
  {author} {\bibfnamefont {H.-C.}\ \bibnamefont {Zhang}}, \bibinfo {author}
  {\bibfnamefont {R.-Z.}\ \bibnamefont {Huang}}, \bibinfo {author}
  {\bibfnamefont {M.}~\bibnamefont {Qin}},\ and\ \bibinfo {author}
  {\bibfnamefont {T.}~\bibnamefont {Xiang}},\ }\bibfield  {title} {\bibinfo
  {title} {Disentangling critical quantum spin chains with clifford circuits},\
  }\href {https://doi.org/10.1103/PhysRevB.111.085121} {\bibfield  {journal}
  {\bibinfo  {journal} {Phys. Rev. B}\ }\textbf {\bibinfo {volume} {111}},\
  \bibinfo {pages} {085121} (\bibinfo {year} {2025})}\BibitemShut {NoStop}%
\bibitem [{\citenamefont {Schollwöck}(2011)}]{Schollwock2011}%
  \BibitemOpen
  \bibfield  {author} {\bibinfo {author} {\bibfnamefont {U.}~\bibnamefont
  {Schollwöck}},\ }\bibfield  {title} {\bibinfo {title} {The density-matrix
  renormalization group in the age of matrix product states},\ }\href
  {https://doi.org/https://doi.org/10.1016/j.aop.2010.09.012} {\bibfield
  {journal} {\bibinfo  {journal} {Ann. Phys.}\ }\textbf {\bibinfo {volume}
  {326}},\ \bibinfo {pages} {96} (\bibinfo {year} {2011})}\BibitemShut
  {NoStop}%
\bibitem [{\citenamefont {Chan}\ \emph {et~al.}(2016)\citenamefont {Chan},
  \citenamefont {Keselman}, \citenamefont {Nakatani}, \citenamefont {Li},\ and\
  \citenamefont {White}}]{ChanWhite2016}%
  \BibitemOpen
  \bibfield  {author} {\bibinfo {author} {\bibfnamefont {G.~K.-L.}\
  \bibnamefont {Chan}}, \bibinfo {author} {\bibfnamefont {A.}~\bibnamefont
  {Keselman}}, \bibinfo {author} {\bibfnamefont {N.}~\bibnamefont {Nakatani}},
  \bibinfo {author} {\bibfnamefont {Z.}~\bibnamefont {Li}},\ and\ \bibinfo
  {author} {\bibfnamefont {S.~R.}\ \bibnamefont {White}},\ }\bibfield  {title}
  {\bibinfo {title} {Matrix product operators, matrix product states, and ab
  initio density matrix renormalization group algorithms},\ }\href
  {https://doi.org/10.1063/1.4955108} {\bibfield  {journal} {\bibinfo
  {journal} {The Journal of Chemical Physics}\ }\textbf {\bibinfo {volume}
  {145}},\ \bibinfo {pages} {014102} (\bibinfo {year} {2016})},\ \Eprint
  {https://arxiv.org/abs/https://pubs.aip.org/aip/jcp/article-pdf/doi/10.1063/1.4955108/14000039/014102\_1\_online.pdf}
  {https://pubs.aip.org/aip/jcp/article-pdf/doi/10.1063/1.4955108/14000039/014102\_1\_online.pdf}
  \BibitemShut {NoStop}%
\bibitem [{\citenamefont {Or{\'u}s}(2014)}]{Orus2014}%
  \BibitemOpen
  \bibfield  {author} {\bibinfo {author} {\bibfnamefont {R.}~\bibnamefont
  {Or{\'u}s}},\ }\bibfield  {title} {\bibinfo {title} {A practical introduction
  to tensor networks: Matrix product states and projected entangled pair
  states},\ }\href@noop {} {\bibfield  {journal} {\bibinfo  {journal} {Annals
  of physics}\ }\textbf {\bibinfo {volume} {349}},\ \bibinfo {pages} {117}
  (\bibinfo {year} {2014})}\BibitemShut {NoStop}%
\bibitem [{\citenamefont {Jordan}\ and\ \citenamefont
  {Wigner}(1928)}]{JordanWigner1928}%
  \BibitemOpen
  \bibfield  {author} {\bibinfo {author} {\bibfnamefont {P.}~\bibnamefont
  {Jordan}}\ and\ \bibinfo {author} {\bibfnamefont {E.}~\bibnamefont
  {Wigner}},\ }\bibfield  {title} {\bibinfo {title} {{{\"U}ber das Paulische
  {\"A}quivalenzverbot}},\ }\href {https://doi.org/10.1007/BF01331938}
  {\bibfield  {journal} {\bibinfo  {journal} {Z. Physik}\ }\textbf {\bibinfo
  {volume} {47}},\ \bibinfo {pages} {631} (\bibinfo {year} {1928})}\BibitemShut
  {NoStop}%
\bibitem [{\citenamefont {Sharma}\ and\ \citenamefont
  {Chan}(2012)}]{SharmaChan2012}%
  \BibitemOpen
  \bibfield  {author} {\bibinfo {author} {\bibfnamefont {S.}~\bibnamefont
  {Sharma}}\ and\ \bibinfo {author} {\bibfnamefont {G.~K.-L.}\ \bibnamefont
  {Chan}},\ }\bibfield  {title} {\bibinfo {title} {Spin-adapted density matrix
  renormalization group algorithms for quantum chemistry},\ }\href
  {https://doi.org/10.1063/1.3695642} {\bibfield  {journal} {\bibinfo
  {journal} {The Journal of Chemical Physics}\ }\textbf {\bibinfo {volume}
  {136}},\ \bibinfo {pages} {124121} (\bibinfo {year} {2012})},\ \Eprint
  {https://arxiv.org/abs/https://pubs.aip.org/aip/jcp/article-pdf/doi/10.1063/1.3695642/13469418/124121\_1\_online.pdf}
  {https://pubs.aip.org/aip/jcp/article-pdf/doi/10.1063/1.3695642/13469418/124121\_1\_online.pdf}
  \BibitemShut {NoStop}%
\bibitem [{\citenamefont {Hubig}\ \emph {et~al.}(2017)\citenamefont {Hubig},
  \citenamefont {McCulloch},\ and\ \citenamefont
  {Schollw\"ock}}]{HubigSchollwock2017}%
  \BibitemOpen
  \bibfield  {author} {\bibinfo {author} {\bibfnamefont {C.}~\bibnamefont
  {Hubig}}, \bibinfo {author} {\bibfnamefont {I.~P.}\ \bibnamefont
  {McCulloch}},\ and\ \bibinfo {author} {\bibfnamefont {U.}~\bibnamefont
  {Schollw\"ock}},\ }\bibfield  {title} {\bibinfo {title} {Generic construction
  of efficient matrix product operators},\ }\href
  {https://doi.org/10.1103/PhysRevB.95.035129} {\bibfield  {journal} {\bibinfo
  {journal} {Phys. Rev. B}\ }\textbf {\bibinfo {volume} {95}},\ \bibinfo
  {pages} {035129} (\bibinfo {year} {2017})}\BibitemShut {NoStop}%
\bibitem [{\citenamefont {White}\ \emph {et~al.}(2018)\citenamefont {White},
  \citenamefont {Zaletel}, \citenamefont {Mong},\ and\ \citenamefont
  {Refael}}]{WhiteRefael2018}%
  \BibitemOpen
  \bibfield  {author} {\bibinfo {author} {\bibfnamefont {C.~D.}\ \bibnamefont
  {White}}, \bibinfo {author} {\bibfnamefont {M.}~\bibnamefont {Zaletel}},
  \bibinfo {author} {\bibfnamefont {R.~S.~K.}\ \bibnamefont {Mong}},\ and\
  \bibinfo {author} {\bibfnamefont {G.}~\bibnamefont {Refael}},\ }\bibfield
  {title} {\bibinfo {title} {Quantum dynamics of thermalizing systems},\ }\href
  {https://doi.org/10.1103/PhysRevB.97.035127} {\bibfield  {journal} {\bibinfo
  {journal} {Phys. Rev. B}\ }\textbf {\bibinfo {volume} {97}},\ \bibinfo
  {pages} {035127} (\bibinfo {year} {2018})}\BibitemShut {NoStop}%
\bibitem [{\citenamefont {Roy}\ and\ \citenamefont
  {Slagle}(2024)}]{RoySlagle2024}%
  \BibitemOpen
  \bibfield  {author} {\bibinfo {author} {\bibfnamefont {S.~G.}\ \bibnamefont
  {Roy}}\ and\ \bibinfo {author} {\bibfnamefont {K.}~\bibnamefont {Slagle}},\
  }\bibfield  {title} {\bibinfo {title} {Reweighted time-evolving block
  decimation for improved quantum dynamics simulations},\ }\href
  {https://arxiv.org/abs/2412.08730} {\bibfield  {journal} {\bibinfo  {journal}
  {arXiv:2412.08730}\ } (\bibinfo {year} {2024})}\BibitemShut {NoStop}%
\bibitem [{\citenamefont {C\'orcoles}\ \emph {et~al.}(2013)\citenamefont
  {C\'orcoles}, \citenamefont {Gambetta}, \citenamefont {Chow}, \citenamefont
  {Smolin}, \citenamefont {Ware}, \citenamefont {Strand}, \citenamefont
  {Plourde},\ and\ \citenamefont {Steffen}}]{CorcolesSteffen2013}%
  \BibitemOpen
  \bibfield  {author} {\bibinfo {author} {\bibfnamefont {A.~D.}\ \bibnamefont
  {C\'orcoles}}, \bibinfo {author} {\bibfnamefont {J.~M.}\ \bibnamefont
  {Gambetta}}, \bibinfo {author} {\bibfnamefont {J.~M.}\ \bibnamefont {Chow}},
  \bibinfo {author} {\bibfnamefont {J.~A.}\ \bibnamefont {Smolin}}, \bibinfo
  {author} {\bibfnamefont {M.}~\bibnamefont {Ware}}, \bibinfo {author}
  {\bibfnamefont {J.}~\bibnamefont {Strand}}, \bibinfo {author} {\bibfnamefont
  {B.~L.~T.}\ \bibnamefont {Plourde}},\ and\ \bibinfo {author} {\bibfnamefont
  {M.}~\bibnamefont {Steffen}},\ }\bibfield  {title} {\bibinfo {title} {Process
  verification of two-qubit quantum gates by randomized benchmarking},\ }\href
  {https://doi.org/10.1103/PhysRevA.87.030301} {\bibfield  {journal} {\bibinfo
  {journal} {Phys. Rev. A}\ }\textbf {\bibinfo {volume} {87}},\ \bibinfo
  {pages} {030301} (\bibinfo {year} {2013})}\BibitemShut {NoStop}%
\bibitem [{\citenamefont {Sun}\ \emph {et~al.}(2018)\citenamefont {Sun},
  \citenamefont {Berkelbach}, \citenamefont {Blunt}, \citenamefont {Booth},
  \citenamefont {Guo}, \citenamefont {Li}, \citenamefont {Liu}, \citenamefont
  {McClain}, \citenamefont {Sayfutyarova}, \citenamefont {Sharma},
  \citenamefont {Wouters},\ and\ \citenamefont {Chan}}]{pyscf}%
  \BibitemOpen
  \bibfield  {author} {\bibinfo {author} {\bibfnamefont {Q.}~\bibnamefont
  {Sun}}, \bibinfo {author} {\bibfnamefont {T.~C.}\ \bibnamefont {Berkelbach}},
  \bibinfo {author} {\bibfnamefont {N.~S.}\ \bibnamefont {Blunt}}, \bibinfo
  {author} {\bibfnamefont {G.~H.}\ \bibnamefont {Booth}}, \bibinfo {author}
  {\bibfnamefont {S.}~\bibnamefont {Guo}}, \bibinfo {author} {\bibfnamefont
  {Z.}~\bibnamefont {Li}}, \bibinfo {author} {\bibfnamefont {J.}~\bibnamefont
  {Liu}}, \bibinfo {author} {\bibfnamefont {J.~D.}\ \bibnamefont {McClain}},
  \bibinfo {author} {\bibfnamefont {E.~R.}\ \bibnamefont {Sayfutyarova}},
  \bibinfo {author} {\bibfnamefont {S.}~\bibnamefont {Sharma}}, \bibinfo
  {author} {\bibfnamefont {S.}~\bibnamefont {Wouters}},\ and\ \bibinfo {author}
  {\bibfnamefont {G.~K.-L.}\ \bibnamefont {Chan}},\ }\bibfield  {title}
  {\bibinfo {title} {Pyscf: the python-based simulations of chemistry
  framework},\ }\href {https://doi.org/https://doi.org/10.1002/wcms.1340}
  {\bibfield  {journal} {\bibinfo  {journal} {WIREs Computational Molecular
  Science}\ }\textbf {\bibinfo {volume} {8}},\ \bibinfo {pages} {e1340}
  (\bibinfo {year} {2018})},\ \Eprint
  {https://arxiv.org/abs/https://wires.onlinelibrary.wiley.com/doi/pdf/10.1002/wcms.1340}
  {https://wires.onlinelibrary.wiley.com/doi/pdf/10.1002/wcms.1340}
  \BibitemShut {NoStop}%
\bibitem [{\citenamefont {Menczer}\ \emph {et~al.}(2024)\citenamefont
  {Menczer}, \citenamefont {van Damme}, \citenamefont {Rask}, \citenamefont
  {Huntington}, \citenamefont {Hammond}, \citenamefont {Xantheas},
  \citenamefont {Ganahl},\ and\ \citenamefont {Legeza}}]{MenczerLegeza2024}%
  \BibitemOpen
  \bibfield  {author} {\bibinfo {author} {\bibfnamefont {A.}~\bibnamefont
  {Menczer}}, \bibinfo {author} {\bibfnamefont {M.}~\bibnamefont {van Damme}},
  \bibinfo {author} {\bibfnamefont {A.}~\bibnamefont {Rask}}, \bibinfo {author}
  {\bibfnamefont {L.}~\bibnamefont {Huntington}}, \bibinfo {author}
  {\bibfnamefont {J.}~\bibnamefont {Hammond}}, \bibinfo {author} {\bibfnamefont
  {S.~S.}\ \bibnamefont {Xantheas}}, \bibinfo {author} {\bibfnamefont
  {M.}~\bibnamefont {Ganahl}},\ and\ \bibinfo {author} {\bibfnamefont
  {O.}~\bibnamefont {Legeza}},\ }\bibfield  {title} {\bibinfo {title} {Parallel
  implementation of the density matrix renormalization group method achieving a
  quarter petaflops performance on a single dgx-h100 gpu node},\ }\href
  {https://doi.org/10.1021/acs.jctc.4c00903} {\bibfield  {journal} {\bibinfo
  {journal} {Journal of Chemical Theory and Computation}\ }\textbf {\bibinfo
  {volume} {20}},\ \bibinfo {pages} {8397} (\bibinfo {year}
  {2024})}\BibitemShut {NoStop}%
\bibitem [{\citenamefont {Menczer}\ \emph {et~al.}(2025)\citenamefont
  {Menczer}, \citenamefont {Werner}, \citenamefont {Hammond}, \citenamefont
  {Xantheas}, \citenamefont {Ganahl}, \citenamefont {Neese} \emph
  {et~al.}}]{MenczerFrank2025}%
  \BibitemOpen
  \bibfield  {author} {\bibinfo {author} {\bibfnamefont {A.}~\bibnamefont
  {Menczer}}, \bibinfo {author} {\bibfnamefont {A.}~\bibnamefont {Werner}},
  \bibinfo {author} {\bibfnamefont {J.}~\bibnamefont {Hammond}}, \bibinfo
  {author} {\bibfnamefont {S.~S.}\ \bibnamefont {Xantheas}}, \bibinfo {author}
  {\bibfnamefont {M.}~\bibnamefont {Ganahl}}, \bibinfo {author} {\bibfnamefont
  {F.}~\bibnamefont {Neese}}, \emph {et~al.},\ }\bibfield  {title} {\bibinfo
  {title} {Orbital optimization of large active spaces via ai-accelerators},\
  }\href {https://arxiv.org/abs/2503.20700} {\bibfield  {journal} {\bibinfo
  {journal} {arXiv:2503.20700}\ } (\bibinfo {year} {2025})}\BibitemShut
  {NoStop}%
\bibitem [{Dat()}]{DataRepo}%
  \BibitemOpen
  \href@noop {} {\bibinfo  {journal}
  {https://github.com/guochu/QCData/tree/master/Clifford\_augmented\_density\_matrix\_renormalization\_group\_for\_ab\_initio\_quantum\_chemistry}\
  }\BibitemShut {NoStop}%
\bibitem [{\citenamefont {Huang}\ \emph {et~al.}(2025)\citenamefont {Huang},
  \citenamefont {Qian}, \citenamefont {Li},\ and\ \citenamefont
  {Qin}}]{HuangQin2025}%
  \BibitemOpen
\bibfield  {journal} {  }\bibfield  {author} {\bibinfo {author} {\bibfnamefont
  {J.}~\bibnamefont {Huang}}, \bibinfo {author} {\bibfnamefont
  {X.}~\bibnamefont {Qian}}, \bibinfo {author} {\bibfnamefont {Z.}~\bibnamefont
  {Li}},\ and\ \bibinfo {author} {\bibfnamefont {M.}~\bibnamefont {Qin}},\
  }\bibfield  {title} {\bibinfo {title} {Augmenting density matrix
  renormalization group with matchgates and clifford circuits},\ }\href
  {https://arxiv.org/abs/2505.08635} {\bibfield  {journal} {\bibinfo  {journal}
  {arXiv:2505.08635}\ } (\bibinfo {year} {2025})}\BibitemShut {NoStop}%
\end{thebibliography}

%

\end{document}